\documentclass[useAMS,usenatbib]{mnras}
\usepackage{graphicx}
\usepackage{amssymb}
\graphicspath{{./image/}}

\usepackage{hyperref}

\newcommand{\HI}{\rm H~{\sc i }}
\newcommand{\HII}{\rm H~{\sc ii }}

\newcommand{\TB}{\delta T_{\rm b}}
\newcommand{\MSUN}{{\rm M}_{\odot}}

\newcommand{\XHI}{x_{\rm HI}}

\newcommand{\TS}{T_{\rm S}}
\newcommand{\TK}{T_{\rm K}}
\newcommand{\TCMB}{T_{\gamma}}
\newcommand{\lya}{\rm {Ly{\alpha}}}

\newcommand{\OmegaB}{\Omega_{\rm B}}
\newcommand{\Omegam}{\Omega_{\rm m}}

\title[Imaging the first sources using SKA]{Imaging the redshifted 21-cm pattern around the first sources during the cosmic dawn using the SKA}

\author[Ghara et.al]{Raghunath Ghara$^1$\thanks{Email: raghunath@ncra.tifr.res.in}, T. Roy Choudhury$^1$, Kanan K. Datta$^{2}$ and Samir Choudhuri$^{3}$ \\
$^1$ National Centre for Radio Astrophysics, TIFR, Post Bag 3, Ganeshkhind, Pune 411007, India\\ 
$^2$ Department of Physics, Presidency University, 86/1 College Street, Kolkata - 700073, India\\
$^3$ Department of Physics, \& Centre for Theoretical Studies, IIT Kharagpur, Kharagpur 721 302, India}

\date{Accepted ?; Received ??; in original form ???}

\pubyear{?}

\begin{document}
\label{firstpage}
\pagerange{\pageref{firstpage}--\pageref{lastpage}}
\maketitle

% Abstract------------------------------------------------------------

\begin{abstract}
Understanding properties of the first sources in the Universe using the redshifted \HI ~21-cm signal is one of the major aims of present and upcoming low-frequency experiments. We investigate the possibility of imaging the redshifted 21-cm pattern around the first sources during the cosmic dawn using the SKA1-low. We model the \HI ~21-cm image maps, appropriate for the SKA1-low, around the first sources consisting of stars and X-ray sources within galaxies. In addition to the system noise, we account also for the astrophysical foregrounds by adding them to the signal maps. We find that after subtracting the foregrounds using a polynomial fit and suppressing the noise by smoothing the maps over $10\arcmin - 30\arcmin$ angular scale, the isolated sources at $z \sim 15$ are detectable with  $\sim 4 - 9 \, \sigma$ confidence level in 2000 h of observation with the SKA1-low. Although the 21-cm profiles around the sources get altered because of the Gaussian smoothing, the images can still be used to extract some of the source properties. We account for overlaps in the patterns of the individual sources by generating realistic \HI ~21-cm maps of the cosmic dawn that are based on $N$-body simulations and a one-dimensional radiative transfer code. We find that these sources should be detectable in the SKA1-low images at $z = 15$ with an SNR of $\sim 14 (4)$ in 2000 (200) h of observations. One possible observational strategy thus could be to observe multiple fields for shorter observation times, identify fields with SNR $\gtrsim 3$ and observe these fields for much longer duration. Such observations are expected to be useful in constraining the parameters related to the first sources.
\end{abstract}

\begin{keywords}
radiative transfer - galaxies: formation - intergalactic medium - cosmology: theory - dark ages, reionization, first stars - X-rays: galaxies
\end{keywords}

%%%%%%%%%%%%%%%

\section{Introduction}
\label{intro}

Detection of the first sources of radiation in the universe which appeared during the ``cosmic dawn'' is at the forefront of modern observational astronomy. It is believed that these sources formed within the dark matter haloes sometime around redshifts $z \sim 15 - 20$ \citep{2007ApJ...665..899W, 2010ApJ...716..510G, 2011ApJ...731...54P, wise2012}. Observing these first sources will not only reveal their unknown properties but also help us in understanding their influence on the formation and evolution of astrophysical objects during later epochs. In recent times a large number of galaxies have been detected at redshift $z \gtrsim 6$ using the broad-band colour \citep{Ellis13, Bouwens15} and the narrow-band $\lya$ emission \citep[e.g.,][]{Ouchi10, Hu10, Kashikawa11}. In addition, a significant number of bright quasars have been detected at high redshifts through various surveys \citep{Fan06b, Venemans15}. New space missions in the near future, e.g., the James Webb Space Telescope ($JWST$)\footnote{http://jwst.nasa.gov}, are expected to detect the most faint sources at even higher redshifts. 

In addition to the above, 21-cm radiation from the neutral hydrogen (\HI) in the intergalactic medium (IGM) can also be used as a probe to detect the very early sources. Motivated by this fact, many of the present low-frequency radio telescopes like the Low Frequency Array (LOFAR)\footnote{http://www.lofar.org/} \citep{van13}, the Precision Array for Probing the Epoch of Reionization (PAPER)\footnote{http://eor.berkeley.edu/} \citep{parsons13}, the Murchison Widefield Array (MWA)\footnote{http://www.mwatelescope.org/} \citep{bowman13, tingay13}, the Giant Metrewave Radio Telescope (GMRT)\footnote{http://www.gmrt.tifr.res.in}\citep{ghosh12, paciga13} etc have dedicated a large amount of their observing resources to detect the signal from the epoch of reionization (EoR). While most of these telescopes are still not able to probe the very early stages of the EoR as they lack the very low-frequency detectors, the future radio telescope like the  Square Kilometre Array (SKA)\footnote{http://www.skatelescope.org/} is expected to detect the signal even from the cosmic dawn. While the first generation telescopes are expected to detect the signal from the EoR statistically (e.g., in terms of the rms, power spectrum, skewness etc), the highly sensitive SKA1-low should be able to image the signal from cosmic \HI ~\citep{2015aska.confE..10M, 2015aska.confE..15W}. 

Recently, many studies have been done using analytical calculations \citep[e.g.,][]{furlanetto04, 2014MNRAS.442.1470P}, semi-numerical simulations \citep{zahn2007, mesinger07, santos08, Thom09, choudhury09, ghara15a, ghara15b}, and full numerical simulations involving radiative transfer \citep{Iliev2006, mellema06, McQuinn2007, shin2008, baek09} to understand the behaviour of the redshifted 21-cm signal from the cosmic dawn and EoR  for different source models.  Though most of these studies have concentrated in detecting the signal using statistical quantities, it will be interesting to study the detectability using imaging techniques. Some recent attempts have been made to understand the detection possibility of large ionized bubbles with LOFAR, MWA, GMRT \citep{kanan2007MNRAS.382..809D, 2008MNRAS.386.1683G, 2008MNRAS.391.1900D, 2009MNRAS.399L.132D, 2011MNRAS.413.1409M, datta2012a, Datta2012b}. In addition, \citet{2012MNRAS.425.2964Z} show that the redshifted 21-cm signal from the EoR can be detected in low-resolution images with LOFAR. Studies have also been done in the same context to detect the signal in post-reionization epochs with SKA1 \citep{2014JCAP...09..050V}.   Our earlier work \citet[][hereafter paper I]{ghara15c} investigated the detectability of very early sources like metal-free Population III (PopIII) stars, galaxies containing Population II (PopII) stars, mini-QSOs and high-mass X-ray binaries (HMXBs) in the presence of system noise and astrophysical foregrounds using a visibility based techniques. The study showed that the SKA1-low should be able to detect the signal from the sources like the PopII stars, mini-QSOs and HMXBs with $\sim 9-\sigma$ confidence by integrating the visibilities signal over all baselines and frequency channels within $\sim 1000$ hours of observation time.

Once the signal from the cosmic dawn is detected, the challenge would be to interpret it and understand the properties of the first sources and the surrounding IGM. One probably needs to use some sophisticated parameter estimation method like the Markov chain Monte Carlo (MCMC) to extract the relevant information. However, before getting involved in the complexities of the parameter estimation methods, one needs to set up appropriate observational strategies to detect the signal.  Detection of the signal from the cosmic dawn is itself very challenging as it is very weak compared to the system noise and the astrophysical foregrounds. In general, one has to integrate the signal over a large observing time to reduce the noise and also use some efficient foreground subtraction method to recover the signal given that the foregrounds are 4-5 orders of magnitude stronger. In this paper, we explore, in detail, the detection of the early sources during the cosmic dawn in \HI ~21-cm images in the presence of system noise and the foregrounds. Our analysis is based on realistic simulations of the signal, system noise, and the relevant astrophysical foregrounds, followed by predictions related to the detectability of the early sources using the SKA1-low. These predictions would be quite useful to plan for observational strategies for detecting the sources in 21-cm observations.

The paper is organized in the following way. In section \ref{simu}, we describe the simulations used in this work. In particular, we describe the model for the sources used in this study in section \ref{source_rt}, while simulating the baseline distribution of the SKA1-low is described in section \ref{ska_base_dist}. The methods to simulate the signal maps, system noise maps and foregrounds maps are described in section \ref{sig_map}, \ref{noi_map} and \ref{FG_map} respectively. The main results of the paper are given in section \ref{res} before we conclude in section \ref{conc}. We choose the Cosmological parameters  $\Omegam=0.32$, $\Omega_\Lambda=0.68$, $\OmegaB=0.049$, $h=0.67$, $n_{\rm s}=0.96$, and $\sigma_8=0.83$, which are consistent with the recent  $Planck$ mission results \citep{Planck2013}.

%%%%%%%%%%%%%%%

\section{SIMULATION}
\label{simu}

The study of detectability of the first sources would require careful modelling of these sources, as well as that of the system noise and the astrophysical foregrounds. We discuss the methods for simulating each of these components in this section.

\subsection{Radiating sources}
\label{source_rt}

\begin{figure*}
\begin{center}
\includegraphics[scale=0.4]{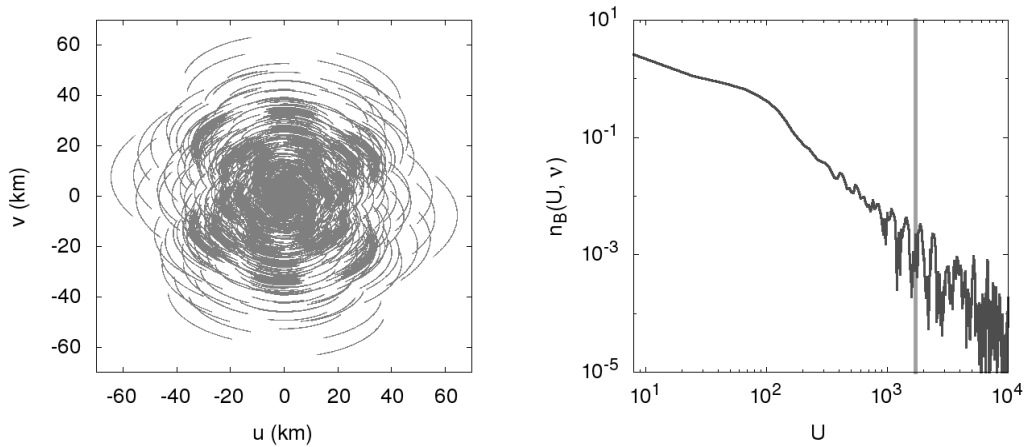}
    \caption{Left-hand panel: The baseline coverage of the SKA1-low for 4 h of observation at a declination $\delta_{\rm dec}=-30^{\circ}$. The integration time taken in this study is 10 sec.  Right-hand panel: The circularly averaged baseline distribution for the SKA1-low at frequency 90 MHz as a function of baseline $U$. The quantity $n_{\rm B}(U,\nu)$ denotes the number density of antenna pairs having baseline $U$ at frequency $\nu$. The vertical line in the right-hand panel represents the baseline corresponding to an angular resolution of $2^{'}$.}
   \label{image_ska_base_alter}
\end{center}
\end{figure*}

The physical conditions of the universe when the first sources formed are relatively poorly understood, and hence the properties of these sources are difficult to model. In this study, we consider different types of sources that could have existed in the early universe, i.e., the PopII stars in the primordial galaxies \citep{2014MNRAS.442.2560W, 2015ApJ...807L..12O, 2016arXiv160407842X}, the mini-QSOs \citep{2003ApJ...596...34B, 2006ApJ...648L...1C, 2007MNRAS.375.1269Z, thomas08, 2009ApJ...701L.133A, 2012MNRAS.425.2974T} and the HMXBs  \citep{2010MNRAS.403...45S, 2011A&A...528A.149M, Fialkov14, 2014MNRAS.440L..26K, 2014MNRAS.445.2034K, 2015ApJ...802....8A}.  Besides these, the metal free PopIII stars are believed to be the another common source during the cosmic dawn. It is however expected that the individual PopIII stars may not be detected in the observations of the redshifted 21-cm signal even with advanced telescopes like the SKA1-low because of the very small region of influence (see Paper I). Thus we have not considered them in this study. While we consider the PopII stars combined with the mini-QSOs in galaxies as our fiducial model source, we will briefly discuss the detectability of other sources too. In the following, we summarize the main properties of the sources used in this paper and refer the reader to Paper I for the details of their spectral energy distribution (SED).

We assume that the first galaxies form in a low-metal region with metallicity $10^{-3} ~ Z_\odot$ \citep{Lai07, Finkelstein09b, 2010MNRAS.407.1003M}  and consist of PopII stars.  We assume that the stars follow a Salpeter IMF with masses between 1 to 100 $\MSUN$. The SED of the stellar component of the model galaxy is generated using the stellar synthesis code {\sc pegase2}\footnote{\tt http://www2.iap.fr/pegase/} \citep{Fioc97}. We assume that a fraction $f_{\rm esc}$ of the UV photons in the intrinsic spectrum of the galaxies escape from the source into the IGM. The SED of the stellar component does not have any significant X-ray photons that are required for heating the neutral / partially ionized IGM. We call this model, consisting of only stars, the {\bf Galaxy model}.

In the {\bf mini-QSO model}, we assume the model source SED has, in addition to the stellar component, a component powered by intermediate mass black holes of mass $10^3-10^6\, \MSUN$. We assume that the X-rays emitted from these mini-QSOs follow a power law with a spectral index $\alpha$ \citep{vanden01, vignali03}. We also introduce a parameter $f_X$ which is the ratio of the X-ray to UV luminosity, where we assume that the UV and X-ray bands span from 10.2 to 100 eV and 100 eV to 10 KeV respectively.  We assume that our fiducial source model has stellar mass  $M_{\star} =  10^7~ \MSUN$, age $t_{\rm age} = 20$ Myr.  We choose $f_{\rm esc}=0.1$, $\alpha=1.5$ \citep{laor97, vanden01, vignali03} and $f_X=0.05$ as the fiducial values.\footnote{The fiducial stellar mass of the source corresponds to stellar content in a dark matter halo of mass $ \sim 6 \times 10^8\, \MSUN$ assuming  $f_{\star} = 0.1$ where $f_{\star}$ is the fraction of baryons converted into stars. In presence of molecular cooling, star formation is possible even in haloes with mass as small as $10^6 ~\MSUN$. However, the efficiency of the formation of molecular hydrogen is very uncertain at high redshift. Detecting the sources formed in such low-mass haloes will be quite challenging \citep[see e.g.,][]{ghara15c}.  In case the star formation occurs only in haloes where the gas cools by atomic transitions, the sources will be hosted in haloes with mass $\gtrsim 10^8 ~\MSUN$, similar to our fiducial value.}  We set the fiducial value of the parameter as $f_X = 0.05$. This corresponds to an accreting BH to galaxy mass ratio of $\sim 10^{-3}$ which is consistent with observations e.g., \citealt{Rix04}. The lifetime of the early sources is very uncertain, though the sources are expected to be short-lived \citep{Meyn05}. Here we set the fiducial age to be $t_{\rm age}=20$ Myr.

The high-mass X-ray binaries could have been another potential source of X-rays in the galaxies. The shape of the SED of the HMXBs depends on the interstellar absorption of the soft X-ray photons and is very uncertain for the high redshift HMXBs. The SED of the HMXBs used throughout the paper is taken from \citet{frag1, frag2}. Due to the significant amount of absorption of the soft X-rays in the interstellar medium of the galaxy, the soft X-ray part of the SED is almost absent from the intrinsic SED of the source. We call this model consisting of stars and HMXBs within galaxies as the {\bf HMXB model}.

%%%%%%%%%%%%%%%

\subsection{Baseline distribution of the SKA1-low}
\label{ska_base_dist}

An important component for simulating radio maps similar to those ones would obtain in observations is the baseline distribution of the telescope. The only telescope considered in this work is the SKA1-low which holds the promise of imaging the high redshift cosmological signal. As per the presently available design, the SKA1-low has a compact core of radius 350 m with closely packed 40 super-stations distributed in four concentric rings, while six closely packed antenna form a super-station. Rest of the 54 super-stations are distributed in a three-arm spiral from 350 m to 35 km radius, where the super-station density distribution follow a logarithmic relation\footnote{The antennae positions for the SKA1-low is taken from http://astronomers.skatelescope.org/wp-content/uploads/2015/11/SKA1-Low-Configuration\textunderscore V4a.pdf}.   The total number of antenna for the SKA1-low is $N_{\rm ant}$ = 564.  Table \ref{tab1} shows the parameters related to the model-observation used in this study.  The left-hand panel of Figure \ref{image_ska_base_alter} shows the baseline $uv$ coverage for 4 h of observation towards a region with declination $\delta_{\rm dec}=-30^{\circ}$ with the SKA1-low . The right-hand panel of Figure \ref{image_ska_base_alter} shows the circularly averaged baseline distribution of the SKA1-low at frequency 90 MHz. The quantity plotted $n_{\rm B}(U,\nu)$ denotes the number density of antenna pairs having baseline $U$ at frequency $\nu$, and is normalized such that $\int n_{\rm B}(U,\nu) ~d^2U = N_{\rm ant}\times (N_{\rm ant}-1)/2$ is the total number of baselines for the SKA1-low. One can easily notice that the baseline distribution is more concentrated at low baseline regions. Note that we have not actually used this circularly averaged baseline distribution in this study, rather we use the true baseline distribution as obtained from the antenna positions. The minimum and maximum baseline for the SKA1-low at redshift 15 turn out to be $\sim 8.5$ and $\sim 19500$ respectively.

The primary field of view (FOV), which depends on the diameter of the individual antenna and observing frequency, is $\sim 5.5^{\circ}$ for the SKA1-low at redshift $z=15$. The maximum angular size ($\theta_{\rm max}$) that can be sampled in the image depends on the minimum baseline considered for the analysis, although the image can be made over the full primary beam. For example, if the minimum baseline $U_{\rm min} \sim 8.5$, then the maximum angular scale $\theta_{\rm max}$ that can be sampled is $6.7^{\circ}$ (which corresponds to a length scale of $1230$ comoving Mpc at redshift $15$). On the other hand, the angular resolution ($\Delta \theta$) of the image depends on the longest baseline considered for the analysis. For example, the SKA1-low should be able to produce images with highest resolution $0.175$\arcmin as its longest baseline is around $U_{\rm max} \sim 19500$ ~at redshift 15. However, the system noise will be much stronger compared to \HI ~21-cm signal if the image is made at this resolution. We, therefore, make images at coarser $2^{'}$ resolution  in order to keep the noise contribution under control. In order to achieve the default images with $2^{'}$ resolution, we consider baselines only up to $U\sim1720$ and discard larger baselines. We note that only a negligible fraction of the total baselines would be discarded in this process as most of the antennae for the SKA1-low are packed at the central region (see right-hand panel of Figure \ref{image_ska_base_alter}). Depending on the values of $\theta_{\rm max}$ and $\Delta \theta$, we generate the signal, noise and foreground maps in a $N_{\rm pixel}\times N_{\rm pixel}$ grid, where $N_{\rm pixel} = \theta_{\rm max} / \Delta \theta$. For example, for $\theta_{\rm max}=$ $6.7^{\circ}$ and $\Delta \theta = $ 2\arcmin  ~we obtain $N_{\rm pixel}=200$.

\begin{table}
\centering
\small
\tabcolsep 3pt
\renewcommand\arraystretch{1.2}
   \begin{tabular}{ll}
\hline
\hline
    Parameters   & Values \\
\hline
\hline
                Redshift ($z$)                                  &          15                  \\
		Central frequency ($\nu_c$) 			&		$88.75$ MHz	\\
		Band width ($B_{\nu}$)				&		16 MHz							\\
		Frequency resolution ($\Delta \nu_c$)		&		100 kHz							\\
    		Observational time	($t_{\rm obs}$)		& 	2000 h					    \\
		System temperature	($T_{\rm sys}$)		&		$60 \times (300 ~\rm MHz / \nu_c)^{2.55} ~\rm K$ \\
		Number of antennae ($N_{\rm ant}$) 		& 	564	\\					
		Effective collecting area ($A_{\rm eff}$)	&		962 $\rm m^2$ \\

\hline
\end{tabular}
\caption[]{The parameters used in this study for a model-observation at redshift $z$ with the SKA1-low.}
\label{tab1}
\end{table} 

%%%%%%%%%%%%%%%

\subsection{Signal maps}
\label{sig_map}

Let us assume that there is an isolated source radiating photons in the neutral and cold IGM. Our first aim is to study the detectability of the 21-cm pattern around such a source. We later study a more complex and realistic model where multiple sources form within a cosmological volume. The main steps to simulate the \HI ~signal maps around an isolated source are as follows:

\begin{itemize}
\item For a given source model, we generate one-dimensional ionization profiles of the hydrogen and helium species and the kinetic temperature around each source by solving the radiative transfer equations \citep{1994MNRAS.269..563F, thomas08}. The main features of the method are described in details in our earlier works \citet{ghara15a, ghara15c}, which is based on \citet{thomas08}.  We assume that the IGM consists of hydrogen and helium of uniform density contrast $\delta$. We assume that the IGM is completely neutral when the source starts to radiate. 

\item We calculate the $\lya$ photon flux by considering the $\lya$ contribution from the continuum spectrum, recombination in the interstellar medium and the secondary ionization due to the X-rays. The $\TB$ profile, as well as the detectability of the source, critically depends on the $\lya$ flux profile as a function of distance from the source. The full $\lya$ radiative transfer simulations are computationally challenging. Hence we have simply assumed that the $\lya$ photon flux reduces as $1/R^2$ with the radial distance $R$. This assumption is consistent with the detailed radiative transfer simulations of, e.g., \citet{2007A&A...474..365S, 2011A&A...532A..97V, 2012MNRAS.426.2380H} at the large scales.

\item It is then straightforward to calculate the coupling coefficients (collisional, $\lya$ coupling and coupling with the CMBR) which are used to generate the  spin temperature ($\TS$) profile.

\item  The differential brightness temperature $\TB (\vec{\theta}, \nu)$ can be expressed as,
\begin{eqnarray}
 \TB (\vec{\theta}, \nu)  \!\!\!\! & = & \!\!\!\! 27 ~ x_{\rm HI} (\mathbf{x}, z) [1+\delta_{\rm B}(\mathbf{x}, z)] \left(\frac{\OmegaB h^2}{0.023}\right) \nonumber\\
&\times& \!\!\!\!\left(\frac{0.15}{\Omegam h^2}\frac{1+z}{10}\right)^{1/2}\left[1-\frac{\TCMB(z)}{T_{\rm S}(\mathbf{x}, z)}\right]\,\rm{mK},
\nonumber \\
\label{brightnessT}
\end{eqnarray}
where $\vec{\theta}$ is the sky direction and $\nu$ is the  frequency corresponding to the observed region. The quantities  $\delta_{\rm B}(\mathbf{x}, z)$, $x_{\rm HI}(\mathbf{x}, z)$ and $\TCMB(z)$ = 2.73 $\times (1+z)$ K denote  the baryonic density contrast, the neutral hydrogen fraction and the CMBR brightness temperature  respectively at the comoving coordinate $\mathbf{x}$ at a redshift $z=1420~{\rm MHz}/\nu -1$. Here $\mathbf{x}$ is related to the sky position $\vec{\theta}$ by the relation $\mathbf{x} = \left\{r(z) \vec{\theta},~ r(z)\right\}$, where $r(z)$ is the comoving distance to $z$. Note that the above expression does not include the effect of the peculiar velocities of the gas in the IGM.  

\item We use the one-dimensional $\TB$ profile to generate the spherically symmetric  $\TB$ map in the simulation box. The comoving length and grid resolution of the simulation box in the angular directions are determined by $\theta_{\rm max}$ and $\Delta \theta$ respectively. The same two quantities along the line of sight are determined by the frequency band width ($B_\nu$) and frequency resolution ($\Delta \nu_c$) of the observation.

\item We generate the two-dimensional baseline distribution (or $uv$ coverage) $n^{i,j}_{\rm B}$ in a $N_{\rm pixel}\times N_{\rm pixel}$ grid for $t_{\rm obs}^{uv}=4$~h of observation at a region with declination $\delta_{\rm dec}=-30^{\circ}$,  while the integration time is taken as $\Delta t_{c}=$10 sec. To incorporate the effect of the empty pixels in the two-dimensional baseline distribution, we first obtain the visibilities of the signal at each $uv$ grid point and then multiply the signal with the baseline ($uv$) sampling function, i.e., zero at $uv$ grids which are empty and unity otherwise.  We then obtain the final image (which is usually known as ``dirty'' image) of the signal by performing a two-dimensional inverse Fourier transform of the visibilities. We note that the \HI ~signal in the ``dirty'' image is hardly distinguishable from the input \HI ~signal. This is due to the fact that $uv$ space is nearly filled and there are not many empty $uv$ grids at small baselines where the \HI ~signal is strong.

\end{itemize}

\begin{figure*}
\begin{center}
\includegraphics[scale=0.7]{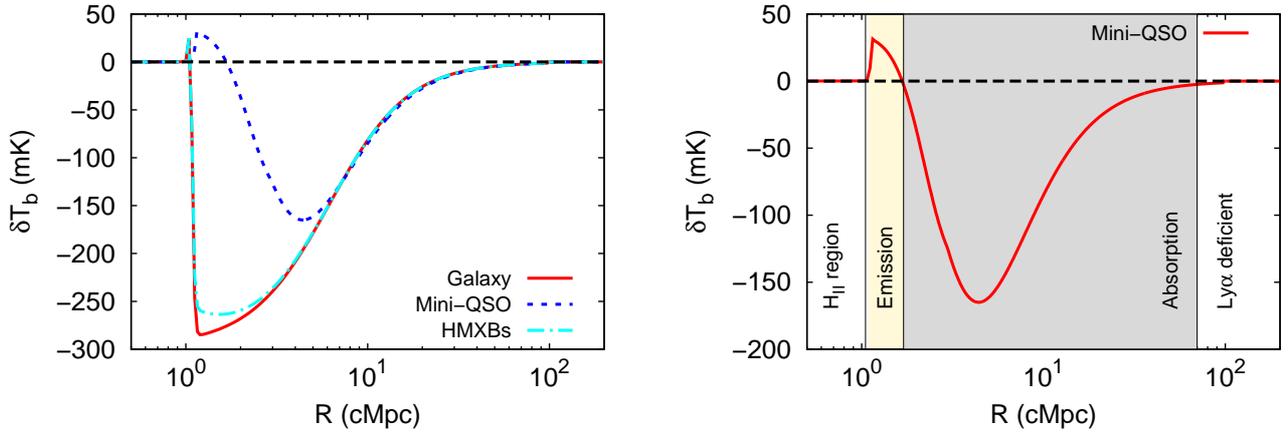}
    \caption{ Left-hand Panel: The radial $\TB$ pattern as a function of the distance $R$ from the centre of the  model source for different source models. The stellar mass of the source is $10^7 ~\MSUN$ for the three models. The ratio of X-ray to UV luminosity of the mini-QSO and HMXB models is 0.05, while the power law spectral index of the mini-QSO model is 1.5. Right-hand panel: Four different regions around the fiducial mini-QSO source model.}
   \label{image_p4spectrum}
\end{center}
\end{figure*}

The left-hand panel of Figure \ref{image_p4spectrum} shows the $\TB$ distribution around the three types of sources considered in this work.   One can easily identify that there exist four separate regions radially outward from the centre of the source (see the right-hand panel of Figure \ref{image_p4spectrum}). These are (1)  \HII ~{\it region}: the signal is zero at the medium just adjacent the source as $\XHI \sim 0$. (2) {\it emission region}: the \HII ~ region is followed by an emission region where $\TS>\TCMB$. (3) {\it absorption region}: the emission region is followed by a strong absorption region where $\TS<\TCMB$ and (4) $\lya$ {\it deficient region}: the signal vanishes at far away region as $\lya$ coupling is not strong enough and thus $\TS = \TCMB$. These results are consistent with earlier works like \citet{2000ApJ...528..597T, 2006ApJ...648...47C, 2006ApJ...648L...1C, 2008ApJ...684...18C, thomas08, Alvarez10, 2014MNRAS.445.3674Y}. The lengths of different regions depend on the source properties. The strength, as well as the volume  of the absorption signal, is much larger than the emission signal for the sources we consider. For example, the minimum $\TB$ for the fiducial mini-QSO model is $\sim -160 $ mK, which is much larger compared to the maximum $\TB$ $\sim$ 30 mK. Thus, one can expect that this region will be the easiest to be detected in radio images, while it may be difficult to identify the \HII ~and emission regions because of the contamination of the weak signal by the system noise and foregrounds. One can also notice that the strength of the absorption signal is larger in the case of the models Galaxy and HMXB compared to the mini-QSO. In other words, one may expect higher detectability for the Galaxy and HMXB source models than the mini-QSO, assuming the sources to be isolated. This will be discussed in more detail in the later part of the paper.

\begin{figure*}
\begin{center}
\includegraphics[scale=0.55]{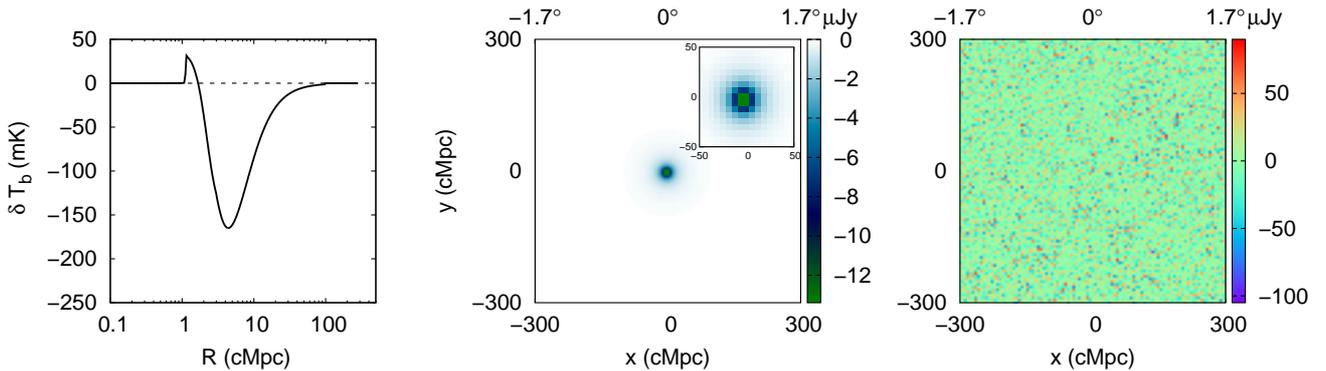}
    \caption{Left-hand panel: The $\TB$ profile of the fiducial source ($M_\star$= $10^7~ \MSUN$, $\delta  = 0$, $\alpha=1.5$, $f_X=0.05$, $t_{\rm age}=20$ Myr) as a function of the radial distance from the centre of the source at redshift 15.  Middle panel: $3.4^{\circ}\times3.4^{\circ}$ image of the signal (without noise) at the frequency channel that contains the centre of the source at redshift 15 for an angular resolution of 2\arcmin. The inner panel represents the zoomed version of the same image. All the color palettes represent $\mu$Jy per beam.  Right-hand panel: The corresponding noise maps at the central frequency channel. The noise map corresponds to a frequency resolution of 100 kHz and 2000 h of observation time.}
   \label{image_p4tbsins_nosmt}
\end{center}
\end{figure*}

The one-dimensional $\TB$ profile around the fiducial mini-QSO source model is shown in the left-hand panel of Figure \ref{image_p4tbsins_nosmt}. The sky specific intensity can be related to $\TB$ as
\begin{equation}
I_{\nu}(\vec{\theta}) = \frac{2 k_B \nu^2  }{c^2} \TB (\vec{\theta}, \nu),
\label{inten}
\end{equation}
where $k_B$ is Boltzmann constant and $c$ is the speed of light. The flux per synthesized beam can be obtained by, 
\begin{equation}
S_{\nu} =I_{\nu}(\vec{\theta}) \times \Delta \Omega,
\label{flux_beam}
\end{equation}
where $\Delta \Omega = (\Delta \theta)^2$ is the beam solid angle. The quantity $S_{\nu}$ thus gives the total flux within a single beam. The middle panel of Figure \ref{image_p4tbsins_nosmt} shows $S_{\nu}$ distribution along the angular directions for our fiducial source for an angular resolution (or beam) 2\arcmin ~at the central frequency channel (which contains the centre of the source in this case). Although the angular extents of our original image are $6.7^{\circ}\times6.7^{\circ}$, we show only a smaller $3.4^{\circ}\times3.4^{\circ}$ image. The maximum amplitude of the signal $S_{\nu}$ in the map is $\sim -13 ~\mu$Jy, with the negative sign signifying that the signal is in absorption.

%%%%%%%%%%%%%%%

\subsection{Noise maps}
\label{noi_map}

The system noise $N(\vec{U}, \nu)$ at different baselines and  frequency channels are uncorrelated and  expected to be Gaussian random variables with zero mean. The rms noise for each  baseline and frequency channel of width $\Delta \nu_c$ and correlator integration time $\Delta t_{c}$ is given by (for single polarization),
\begin{equation}
\sqrt{\left< N^2 \right>} = \frac{\sqrt{2} k_B T_{\rm sys}}{A_{\rm eff} \sqrt{\Delta \nu_{c} ~ \Delta t_{c}}},
\label{rms_noise}
\end{equation}
where $A_{\rm eff}$ is the effective collecting area of each antenna and $T_{\rm sys}$ is the system temperature. Here we have chosen $\Delta t_{c}=10$ sec. The steps to generate the noise maps are given below:

\begin{itemize}

\item First, we generate Gaussian random noise (both the real and imaginary parts) with zero mean and rms $\sqrt{\left< N^2 \right>}$ in the $N_{\rm pixel}\times N_{\rm pixel}$ grid in the Fourier space.

\item  The presence of multiple baselines in a $uv$ grid point can be used to decrease the noise in that pixel. We account for this by simply scaling the noise in $(i,j)$th pixel by a factor $1/\sqrt{n^{i,j}_{\rm B}}$.\footnote{In principle, the baseline distribution is dependent on the frequency of interest and thus, should be different for different frequency channels. In this study, we have ignored this fact and worked with only one baseline distribution which corresponds to the central frequency of the observation.}

\item  By averaging over long observation time $t_{\rm obs}$, one can decrease the noise further by a factor of $\sqrt{t_{\rm obs}/t_{\rm obs}^{uv}}$, which is done in this work as well.

\item   As mentioned earlier, the presence of empty pixels in the two-dimensional baseline distribution is accounted for by including a mask which is zero at the empty pixels and unity otherwise.

\item We obtain the real space noise map by doing two-dimensional inverse Fourier transform of the reduced noise in Fourier space at each frequency channel.

\end{itemize}

The right-hand panel of Figure \ref{image_p4tbsins_nosmt} shows the simulated noise map at the central frequency channel for an angular resolution 2\arcmin ~for 2000 h of observation time and the parameters listed in Table \ref{tab1}.  The rms noise per beam of the corresponding map is $\sim$ 19 $\mu$Jy. The amplitude of the signal as shown in the middle panel of Figure \ref{image_p4tbsins_nosmt} is significantly smaller the rms noise for 2000 h of observation, thus the signal is not detectable without further reducing the noise using some other techniques like ``smoothing'', which we will discuss later part of the paper (in section \ref{RES_SI_NS_FG}).

%%%%%%%%%%%%%%%

\subsection{Foreground maps}
\label{FG_map}

The cosmological signal will be contaminated by other astrophysical foregrounds which have significantly larger amplitude \citep{2008MNRAS.385.2166A, ghosh12}. In this study, we consider the Galactic synchrotron radiation and emission from unresolved extragalactic point sources as the major contributors to these foregrounds. Among these two, the Galactic synchrotron radiation contributes $\sim 70\%$  of the total foregrounds \citep{2006ApJ...650..529W, jelic08}. In addition to these, the Galactic free-free emission, supernova remnants, and extragalactic radio clusters may also provide a small contribution to the total foreground, which has been neglected in this study. The method of simulating the foregrounds is given below:

\begin{itemize}

\item 
{\it Galactic synchrotron radiation:} We have mainly followed \citet{Choudhuri2014MNRAS.445.4351C} for simulating the Galactic synchrotron radiation. The fluctuations in the foregrounds are assumed to be Gaussian random fields with an angular power spectrum $C^{\rm syn}_{2\pi U}(\nu)$ that can be expressed as \citep[see, e.g.,][]{ghosh12},
\begin{equation}
C^{\rm syn}_l(\nu)=A_{150}~ \left( \frac{1000}{l}\right)^{\bar{\beta}} ~\left( \frac{\nu}{\nu_\star} \right)^{-2{\bar{\alpha}_{\rm syn}}-2{\bar{\Delta {\alpha}_{\rm syn}}\log(\frac{\nu}{{\nu}_{\star}})}},
\label{cl_FG}
\end{equation} 
where $\nu_\star = 150$ MHz,  $A_{150} = 513 ~\rm mK^2$, ${\bar{\beta}}=2.34$, ${\bar{\alpha}_{\rm syn}}=2.8$ and ${\bar{\Delta{\alpha}_{\rm syn}}}=0.1$. The parameters for the Galactic synchrotron emission have been taken from \citet{1998ApJ...505..473P, 2006ApJ...650..529W}.

Given the angular power spectrum, we first generate the maps of the temperature fluctuations for the foregrounds in the Fourier space using the relation 
\begin{equation}
\Delta T_{\rm syn}(U, \nu)=\sqrt{\frac{\Omega C^{\rm syn}_{l}(\nu)}{2}} \left[x(U) + iy(U) \right],
\label{FG_t_uv}
\end{equation} 
where $l=2\pi U$ and $\Omega$ is the total solid angle of the simulated area. The quantities $x(U)$ and $y(U)$ are two independent Gaussian random variables with zero mean and unit variance. We then carry out a two-dimensional inverse Fourier transform on the $\Delta T_{\rm syn}(U, \nu)$ distribution to obtain the real space distribution $\delta T_{\rm syn}(\vec{\theta}, \nu)$ at each frequency channel. The specific intensity fluctuation can be simply obtained as $\delta I_{\rm syn}(\vec{\theta}, \nu)= (2 k_B/{\lambda}^2) \delta T_{\rm syn}(\vec{\theta}, \nu)$. We multiply this with the beam solid angle to obtain the flux per synthesized beam for the Galactic synchrotron radiation maps.

\item 
{\it Extragalactic point sources:} The method used to simulate the foregrounds from the extragalactic point sources is based on the observations of \citet{ghosh12} with GMRT at frequency $\nu_\star = 150$ MHz.\footnote{The foreground contribution from the  unresolved extragalactic point sources can be divided into two parts, ($i$) the Poisson contribution and ($ii$) the clustering contribution.  The point source clustering part dominates over the Poisson part at large angular scales \citep{2002ApJ...564..576D}. However, the diffuse synchrotron emission from our galaxy is expected to be much stronger than the point source clustering contribution at these large scales \citep{2002ApJ...564..576D, 2005ApJ...625..575S, kanan2007MNRAS.382..809D}. We, therefore, do not consider the foreground contribution from the clustering part in this study.} The differential source count is given by
\begin{equation}
\frac{dN}{dS}=\frac{10^{3.75}}{\rm Jy.Sr} \left( \frac{S}{\rm Jy}\right)^{-1.6}.
\label{equ_dnds}
\end{equation}
We assume that the point sources with flux larger than $5\sigma$ can be identified and removed from the pixel. In this work, we generate the map for the unresolved extragalactic point sources within a flux range $10^{-4}$ to 0.1 mJy, while we assume that all source above $5\sigma \sim 0.1$ mJy will be resolved and removed. First, we divide the flux range into multiple flux bins and calculate the number of sources associated with each flux bins. We randomly distribute the sources in the two-dimensional map along the angular directions at the central frequency channel. The maps at other frequency channels are generated assuming the flux of the sources changes with frequency as,

\begin{equation}
S_{\nu}=S_{{\nu}_{\star}} \left( \frac{\nu}{{\nu}_{\star}} \right)^{-{\alpha}_{\rm ps}},
\label{eq_pt_flux}
\end{equation}
where ${\alpha}_{\rm ps}$ is the spectral index of the foregrounds contribution from the point sources. We generate the index ${\alpha}_{\rm ps}$ for each point source from a uniform random distribution with values in the range of 0.7 to 0.8.

\begin{figure}
\begin{center}
\includegraphics[scale=0.7]{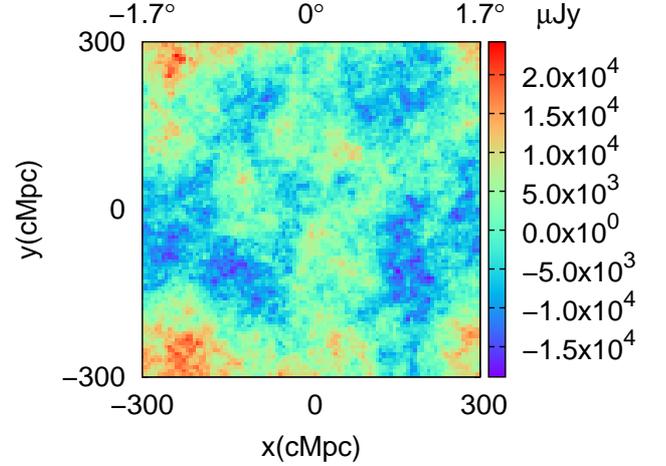}
    \caption{The foreground map at the central frequency channel $\nu_c=88.75$ MHz. The angular resolution of the map is 2\arcmin. The map includes the contributions from the Galactic synchrotron radiation and extragalactic radio emission from the unresolved point sources.}
   \label{image_p4FG}
\end{center}
\end{figure}

\end{itemize}

Figure \ref{image_p4FG} shows the 2\arcmin ~resolution map for the total foreground signal at the central frequency channel with the two-dimensional mean is subtracted out. The strength of the foreground signal is $\sim 3-4$ order of magnitude stronger than the expected signal at this resolution as can be seen by comparing with Figure \ref{image_p4tbsins_nosmt}. Thus, it is obvious that the recovery of the signal in the presence of such a strong foregrounds is indeed a challenging task. We will discuss various techniques to subtract the foregrounds below the signal level at the later part of the paper.
%%%%%%%%%%%%%%%

\section{Results }
\label{res}

In order to estimate the detectability of the first sources through imaging the high redshift 21-cm signal, we choose our fiducial source model to be the mini-QSO. We first study in great detail the simplistic situation where there is a single isolated source in the FOV, and then consider a more realistic situation where there are multiple sources in the field.

\subsection{Isolated source}
\label{res_isolated}

We assume that the isolated source is completely isolated and situated at the centre of the FOV. The fiducial parameters of the mini-QSO model, as mentioned earlier, are taken to be $M_\star = 10^7 ~ \MSUN$, $f_{\rm esc}=0.1$, $f_X=0.05$, $\alpha=$1.5 and age $t_{\rm age}=20$ Myr. We assume the IGM density contrast $\delta=0$. We choose the fiducial angular resolution for imaging as 2\arcmin. The spatial length scale corresponding to this resolution is $\sim 6$ cMpc, which is similar to the radial distance to the strongest absorption signal around the fiducial source.

\subsubsection{Signal and the system noise}
\label{RES_SI_NS}

First, let us consider a scenario where we can ignore the complexities arising from the foregrounds. Even in this simple case where we deal only with the signal and the system noise, we find that the noise is much larger than the cosmological signal as shown earlier in Figure \ref{image_p4tbsins_nosmt}.

One method of increasing the SNR is by smoothing the maps using some filter. We have seen in Paper I that the signal exceeds the system noise only for baselines $U \lesssim 100$, which corresponds to angular scales $\gtrsim 10 \arcmin$. In order to see similar effects in the image, we have used two-dimensional Gaussian filters of different widths (i.e., standard deviations) for smoothing the images at all the frequency channels. These Gaussian filters are applied along the two angular directions at each frequency channel. We have used a fixed frequency width of 100 kHz throughout the paper. In principle, the signal to noise ratio can be increased by an additional smoothing along the frequency direction. However, this will introduce an additional complexity in measuring the brightness temperature profile from a source. The signal evolves along the frequency direction and the evolution is particularly strong around sources that bright in UV and X-ray bands \citep{2012MNRAS.426.3178M}. Using higher frequency width will smooth out this evolution to some extent. Additionally, observational parameters such as the sky temperature and effective antenna area have strong frequency dependencies.  In order to properly calculate the system noise one should consider these effects and we defer this for future work. To avoid all these complexities, we restrict our analysis within a very small frequency interval.

\begin{figure*}
\begin{center}
\includegraphics[scale=0.55]{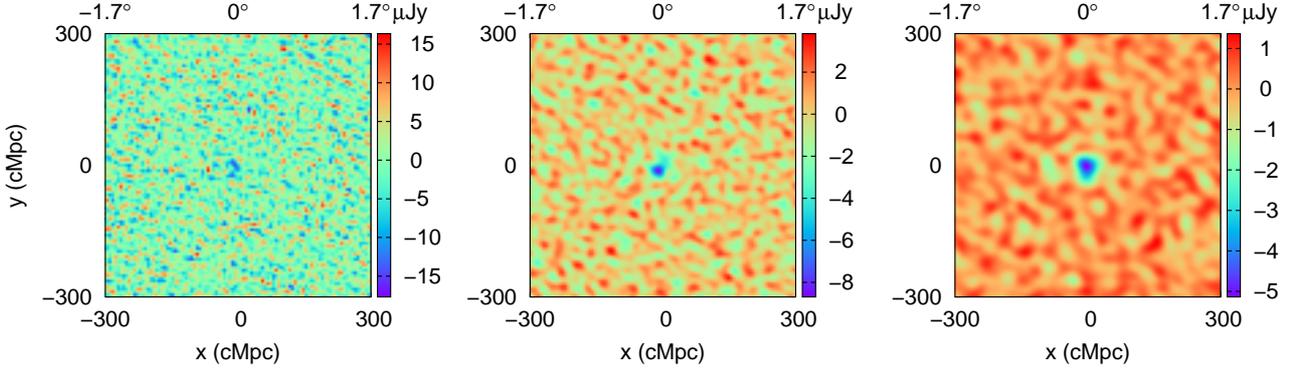}
    \caption{Left-hand to right-hand panels show the images smoothed with a Gaussian kernel of size 10\arcmin, 20\arcmin ~and 30\arcmin  ~respectively. The images contain the signal from the radiating source as well as the system noise. Our fiducial source parameters are  $M_\star$= $10^7~ \MSUN$, $\alpha=1.5$, $f_X=0.05$, $t_{\rm age}=20$ Myr at redshift 15. We have taken 100 kHz frequency resolution and 2000 h of observation time.}
   \label{image_p4diff_kernal}
\end{center}
\end{figure*}

The effect of smoothing on the image maps is shown in Figure \ref{image_p4diff_kernal}. The three panels from the left-hand side show the effect of using a Gaussian smoothing kernel of width $10 \arcmin$, $20 \arcmin$ and $30 \arcmin$ respectively. One can clearly see that the 21-cm pattern of the source becomes more prominent as we increase the width of the kernel. This is related to the fact that the noise amplitude decreases because of smoothing. For example, the rms noise of the map without smoothing (right-hand panel of Figure \ref{image_p4tbsins_nosmt}) is  $\sim$19 $\mu$Jy for 2000 h of observation and 100 kHz of frequency resolution, while the rms noise reduces to $\sim 0.4 ~\mu$Jy for smoothing with the fiducial Gaussian filter of size 30\arcmin ~(right-hand panel of Figure \ref{image_p4tbsins_nosmt}).\footnote{$30^{'}$ corresponds to a spatial scale of $\sim 90$ cMpc, which is typical size of the 21-cm region around the source after 20 Myr.} 

We define the SNR of the maps as the ratio of the largest absolute amplitude of the observed pixel (signal + noise in this case) in the image plane and the rms noise. We average over 10 independent realizations of the noise while calculating the SNR. The SNRs in the left-hand to right-hand panels of Figure \ref{image_p4diff_kernal} are 4, 7.5 and 11 respectively, while the corresponding rms noise are 4.4, 1.1 and 0.4 $\mu$Jy respectively.  It is clear that the SNR increases with the width of the Gaussian filter. Thus, it is possible to detect the signal by smoothing the maps with sufficiently wide filters like 30\arcmin ~within 400 (150) hours of observation  with an SNR $\sim 5$ ($\sim 3$).

We use the Pearson-cross-correlation to quantify the similarity between two maps. For two maps having amplitudes $x_{i}$ and $y_{i}$ at the $i$th pixel, the Pearson-cross-correlation coefficient is defined as
\begin{equation}
\chi=\frac{\sum_{i}{(x_i-\bar{x})(y_i-\bar{y})}}{\sqrt{\sum_{i}(x_i-\bar{x})^2}\sqrt{\sum_{i}(y_i-\bar{y})^2}},
\end{equation}
where $\bar{x}$ and $\bar{y}$ are the mean of the maps $x_i$ and $y_i$ respectively. The value of $\chi$ for the 21-cm map around the isolated fiducial source smoothed by a Gaussian kernel of width $30 \arcmin$ and a similar smoothed map which included the system noise is 0.56. The value of $\chi$ is relatively smaller in these case as the signal from the source is localized in the central part of the map, while most part of the image in the second case is dominated by the noise.

\begin{figure*}
\begin{center}
\includegraphics[scale=0.45]{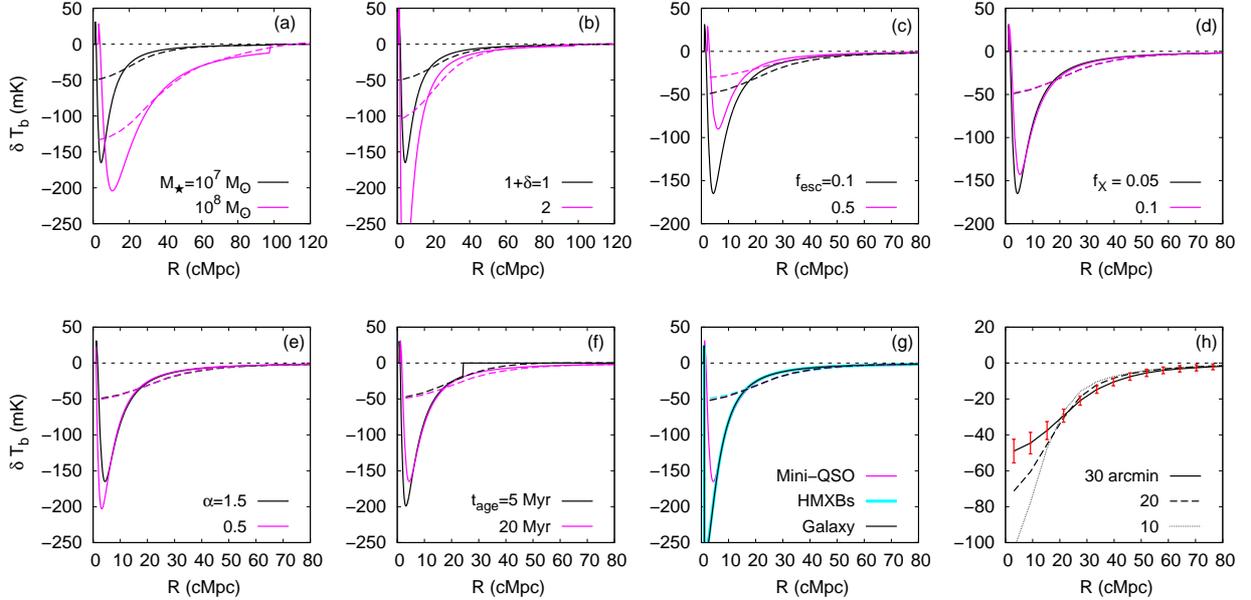}
    \caption{The radial brightness temperature profiles around an isolated source at $z = 15$. The panels (a) to (f) show the dependence of the profiles on the model parameters, namely, the stellar mass $M_{\star}$, the overdensity of the surrounding IGM ($1+\delta$), the UV escape fraction $f_{\rm esc}$, the ratio of X-ray to UV luminosity $f_X$, the X-ray spectral index $\alpha$ and the age of the source $t_{\rm age}$ respectively. The solid curves are for the case without any smoothing, while the corresponding dashed curves represent the corresponding smoothed $\TB$ profile, where the smoothing is done with a  Gaussian filter of width 30\arcmin.  While varying one parameter, we have fixed the other parameters to their fiducial values for generating the $\TB$ profiles.  Panel (g) shows the intrinsic and the smoothed $\TB$ profiles for the three different types of source models considered in this paper. Panel (h) shows the smoothed $\TB$ profiles around the fiducial source for three different widths of Gaussian kernel, namely 30\arcmin, 20\arcmin ~and 10\arcmin. The error bars represent the $1\sigma$ rms of the system noise corresponding to a frequency resolution 100 kHz, 2000 h of observation time and 30\arcmin ~Gaussian filter.}
   \label{image_p4tbprof_smt_allpara}
\end{center}
\end{figure*}

We have seen that the detectability of the signal from the fiducial source improves significantly when we smooth the image over some scale. However, this same smoothing can change the original profile of the $\TB$ distribution around the fiducial source. This may create additional difficulties in extracting the properties of the source from these maps. Thus, we must check whether these smoothed profiles can even be used for parameter estimation. The panels (a) - (h) in Figure \ref{image_p4tbprof_smt_allpara} show the true $\TB$ profiles (solid lines) and the smoothed ones with a Gaussian filter of size 30\arcmin (dashed lines) for different source parameters. In each of these panels, we keep all the parameters except one fixed to their fiducial values. One can easily notice that the smoothed profiles are quite sensitive to parameters like the stellar mass $M_{\star}$, over-density ($1+\delta$) and the UV escape fraction  $f_{\rm esc}$, while the profiles are almost unaffected while changing the X-ray parameters $f_X$, $\alpha$  and the age of the source $t_{\rm age}$. We can thus infer that it should be possible to infer the values of the $M_{\star}, \delta$ and $f_{\rm esc}$ from the smoothed images, while other parameters may remain undetermined. The panel (g) in Figure \ref{image_p4tbprof_smt_allpara} shows the smoothed  $\TB$ profiles for different source models. It is interesting to note that the profiles look almost the same, thus implying that it would not be straightforward to infer the precise source model from the smoothed image maps. The panel (h) of Figure \ref{image_p4tbprof_smt_allpara} shows the smoothed $\TB$ profile of the fiducial source for the Gaussian filters of width 10\arcmin, 20\arcmin ~and 30\arcmin.  The error bars in the panel show the $1-\sigma$ error due to the system noise for 2000 h of observation and 100 kHz of frequency resolution when the filter width is taken to be 30\arcmin. The errors have been obtained by averaging over pixels lying in circular annulus around the centre of the source. The system noise, when averaged circularly in a single slice, should drop like $R^{1/2}$ in a scenario when the noise in adjacent pixels in the image are uncorrelated. However, smoothing makes the noise at different pixels correlated and therefore a simple drop of the error like $R^{1/2}$ is not applicable in this case. We calculate the true error bars by making independent realizations of the noise map and smoothing it using the Gaussian filter. We then bin the image in the radial direction around the centre of the source and calculate the circularly averaged noise at each bin for each realization. The variation of this quantity across realizations gives the required rms. The error bars in panel (h) of Figure \ref{image_p4tbprof_smt_allpara} represent the rms calculated using this method. By comparing with the panels (a) - (g), we find that the change in the profiles when we change the values of $M_{\star}, \delta$ and $f_{\rm esc}$ is larger than the sizes of the error bars. Thus one expects that a reasonably advanced parameter estimation method (e.g., MCMC) using the smoothed $\TB$ profile can put strong constraints on the three parameters $M_{\star}$, ($1+\delta$) and  $f_{\rm esc}$, whereas the X-ray parameters and  $t_{\rm age}$ may not be strongly constrained.

Now let us discuss the detectability of other source models in this foreground-free scenario. The SNRs for the Galaxy and HMXB source models for the fiducial parameter values are 11.3 and 11.2 respectively for the smoothed maps. Although the absorption signal in the expected $\TB$ profiles in Figure \ref{image_p4spectrum} is stronger for the Galaxy and HMXB source models compared to the mini-QSO, all the profiles look almost similar after smoothing which leads to similar SNRs (see panel (g) of Figure \ref{image_p4tbprof_smt_allpara}).  The SNR is also quite sensitive to the redshift of observation. For example, if the source formation starts at a lower redshift, say, $z=10$, the SNR of the fiducial mini-QSO model increases from $\sim$11 to $\sim 46$ because of the decrease in the system noise.  The SNRs for different values of the parameters are listed in Table \ref{table_snr}.

Till now we have been considering the scenario where there is only one source in the FOV and the $\TB$ profile is calculated using the method given in section \ref{sig_map}. Since a small amount of $\lya$ radiation is sufficient to couple $\TS$ to $\TK$, it is possible that the IGM may rapidly attain a state where the $\lya$ coupling is very strong in every part of the IGM. This can arise, e.g., from the significant overlap between the individual $\lya$ bubbles of the very early sources. In such a scenario, we will have $\TS=\TK$ at all points in the IGM which we call the ``$\lya$ coupled scenario'' (same as model $B$ in Paper I). In this case, a large fraction of the IGM show strong absorption signal, however, the mean subtracted signal is expected to be very little in the emission and absorption regions. In order to achieve an SNR of $\sim 5$ for the fiducial source in this scenario, we require an observing time as large as 10,000 h when the smoothing is done with a Gaussian filter of size 30\arcmin. Hence the detectability of the signal will be significantly more challenging when the $\lya$ coupling  complete.

\begin{table*}
\centering
\begin{tabular}{|l|c|c|c|c|c|c|c|c|c|c|c|}
\hline
Source & $M_{\star}$ & $1+\delta$ & $f_{\rm esc}$  & $1+z$ & Filter & SNR1 & SNR2\\
\hline
\hline
Mini-QSO & $10^7 \MSUN$ & 1 & 0.1  & 16 & $30^{'}$         & 11.1    & 9.1 \\
\hline
Mini-QSO & $10^6 \MSUN$ & 1 & 0.1  & 16 & $30^{'}$         & 3.6    & 3.4 \\
\hline
Mini-QSO & $10^8 \MSUN$ & 1 & 0.1  & 16 & $30^{'}$         & 25.9    & 20.2 \\
\hline
Mini-QSO & $10^7 \MSUN$ & 2 & 0.1  & 16 & $30^{'}$         & 20.4    & 17.5 \\
\hline
Mini-QSO & $10^7 \MSUN$ & 1 & 0.5  & 16 & $30^{'}$         & 5.2    & 4 \\
\hline
Mini-QSO & $10^7 \MSUN$ & 1 & 0.1  & 11 & $30^{'}$         & 46    & 38 \\
\hline
Mini-QSO & $10^7 \MSUN$ & 1 & 0.1  & 16 & $10^{'}$         & 4.2    & 4.0 \\
\hline
Galaxy & $10^7 \MSUN$   & 1 & 0.1  & 16 & $30^{'}$         & 11.3    & 9.4 \\
\hline
HMXBs & $10^7 \MSUN$    & 1 & 0.1  & 16 & $30^{'}$         & 11.2    & 9.2 \\
\hline
\end{tabular}
\caption[]{The SNRs for different scenarios considered in the paper. These correspond to an observation time of 2000 h with the 564 antennae SKA1-low with a frequency resolution of 100 kHz. The quantities SNR1 and SNR2 represent the signal to noise ratios for the scenarios with and without foregrounds respectively. }
\label{table_snr}
\end{table*}

%%%%%%%%%%%%%%%

\subsubsection{Signal + Noise + Foregrounds}
\label{RES_SI_NS_FG}

\begin{figure*}
\begin{center}
\includegraphics[scale=0.65]{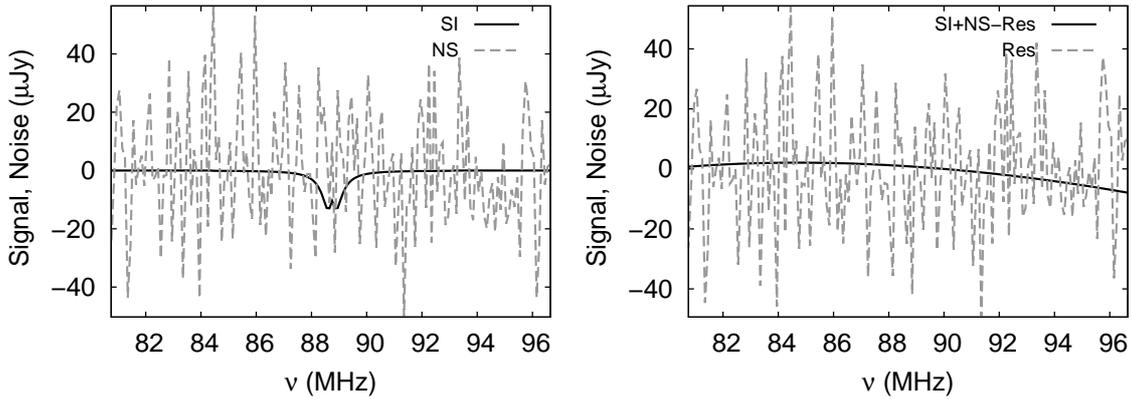}
    \caption{Left-hand panel: The real space signal (solid curve) and noise (dashed curve)  as a function of the frequency channels along the line of sight to the centre of the source.  Right-hand panel: The solid curve represents the difference between the signal + noise (before  the foreground subtraction) and the residual signal + noise (after the foregrounds are subtracted) along the $\nu$ direction. The dashed curve represents the residual signal + noise along the frequency direction after the foregrounds are subtracted using a third order polynomial fitting method.}
   \label{image_p4si_ns_fg}
\end{center}
\end{figure*}

\begin{figure*}
\begin{center}
\includegraphics[scale=0.65]{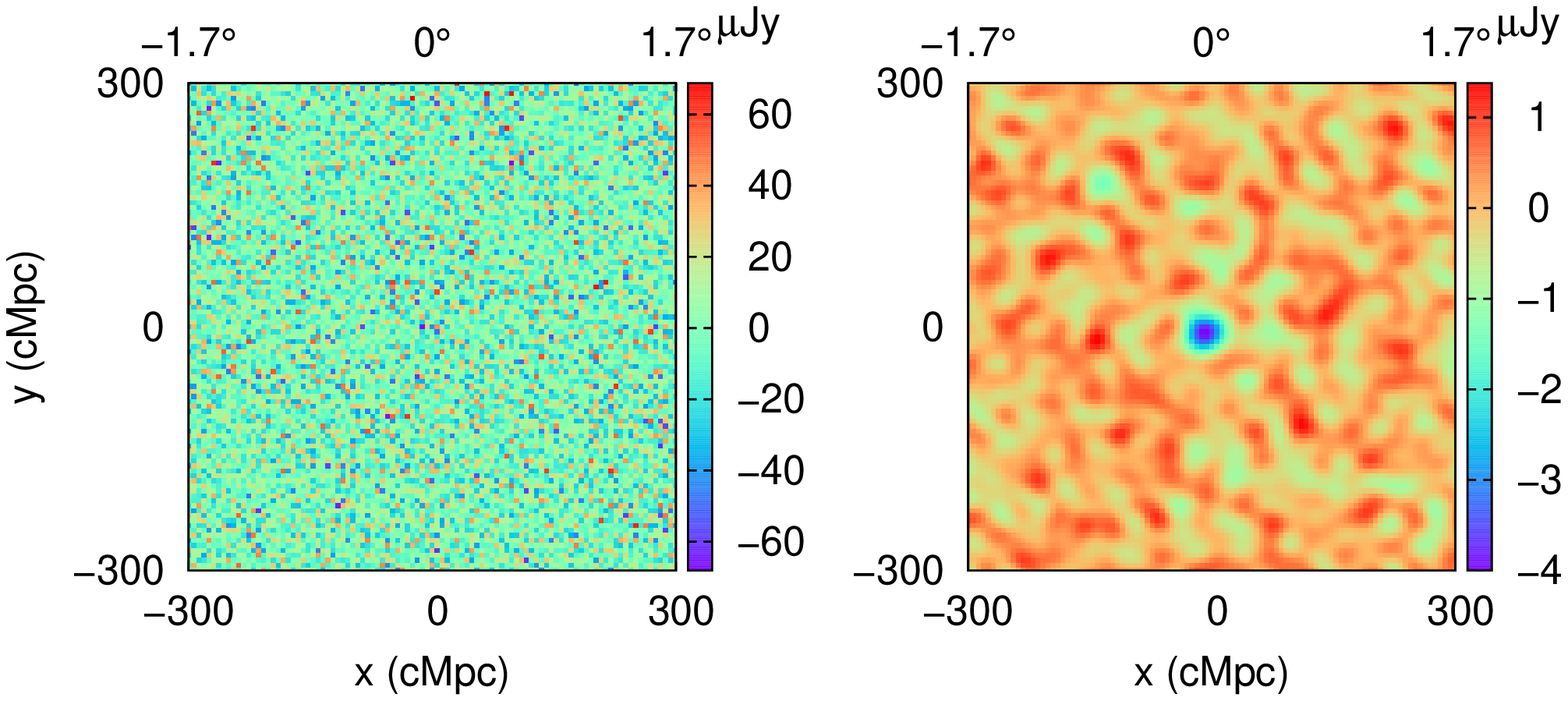}
    \caption{ Left-hand panel : Map of the residual signal and noise after the foregrounds are subtracted (without smoothing). The angular resolution of the map is 2\arcmin. The noise corresponds to 2000 h of observation, 100 kHz of frequency resolution and baseline distribution of 564 antennae SKA1-low. Right-hand panel: Same as the left-hand panel but smoothed with a Gaussian filter of size 30\arcmin.}
   \label{image_p4res_sisnsfg}
\end{center}
\end{figure*}

Let us now investigate the detectability of the first sources in the presence of astrophysical foregrounds. As we have seen that the foregrounds are several orders larger than the signal as well as the system noise, it is in principle a very challenging task to recover the signal. However, the frequency dependence of the foregrounds is relatively smooth, while other components namely the signal and the noise behave differently. This property of the foregrounds can be used to subtract the foregrounds and recover the signal. 

There are many approaches considered for removing the foregrounds, such as  the polynomial fitting based method \citep{2006ApJ...653..815M, 2006ApJ...650..529W, 2008MNRAS.391..383G, jelic08, 2011MNRAS.418.2584G, 2011MNRAS.413.2103P, 2012A&A...540A.129A, 2015MNRAS.447..400A}, Wp smoothing \citep{2010MNRAS.405.2492H}, independent component analysis \citep{2013MNRAS.429..165C}, continuous wavelet transform \citep{2013ApJ...773...38G} and so on. In this work, we consider the polynomial fitting method which is relatively straightforward to implement among the existing ones. The steps we follow to generate the foregrounds subtracted smoothed images are:

\begin{itemize}

\item First, we calculate  the total visibility $V(\vec{U}, \nu)$ which can be written as, 
\begin{equation}
V(\vec{U}, \nu)=S(\vec{U}, \nu) + N(\vec{U}, \nu) + F(\vec{U}, \nu),
\label{equ_visi}
\end{equation}
where $S(\vec{U}, \nu)$, $N(\vec{U}, \nu)$ and $F(\vec{U}, \nu)$ are the contributions from the cosmological signal, system noise and foregrounds respectively.

\item We choose the components of $V(\vec{U}, \nu)$ along the frequency direction for each $\vec{U}$ and separately fit the real and imaginary part using a third order polynomial in logarithmic space. The form of the polynomial is given by
\begin{equation}
\log V(\vec{U}, \nu)=\sum_{m=0}^{n} a_m \left( \log \nu \right)^m ,
\label{equ_polfit}
\end{equation}
where we choose $n=3$ in this case. One thing to remember is that  certain amount of signal and system noise is also removed during the foreground removal process. Thus fitting with a polynomial of a very high order may not be helpful. 

\item After the polynomial fitting, we subtract the fitted visibilities from the total visibilities to obtain the residual visibilities $V_{\rm res}(\vec{U}, \nu)$. These residual visibilities contain the residual foregrounds, signal and noise.  

\item Finally we carry out the two-dimensional inverse Fourier transform of the $V_{\rm res}(\vec{U}, \nu)$ at each frequency channel to get the real space maps that have the foregrounds subtracted. 

\item We smooth the image with a two-dimensional Gaussian filter to reduce the rms noise.

\end{itemize}

The left-hand panel of Figure \ref{image_p4si_ns_fg} shows the real space 21-cm signal and system noise of the SKA1-low for 2000 h of observation as a function of the frequency channels along the line of sight which contains the centre of the source.   One can see that the signal is contaminated by the system noise. The foregrounds along the same line of sight are several orders larger than the signal or the noise and hence is not shown in the figure. The right-hand panel shows the residual signal + noise after subtracting the foreground using the third order polynomial. Also shown is the difference between the original and the residual signal + noise. Clearly, the subtraction method is accurate enough to recover almost the original signal and noise. Figure \ref{image_p4res_sisnsfg} shows the foreground subtracted image without (left-hand panel) and with (right-hand panel) smoothing with the Gaussian filter of size 30\arcmin. One can see that some amount of signal and noise also subtracted during the foregrounds removal process by comparing the images at the right-hand panel of Figure \ref{image_p4res_sisnsfg} and  the right-hand panel of Figure \ref{image_p4diff_kernal}. The loss of signal during the foregrounds removal is consistent with previous works like \citet{jelic08, 2009MNRAS.398..401L, 2010MNRAS.405.2492H, 2011MNRAS.413.2103P, 2012MNRAS.425.2964Z}. The value of $\chi$ for these two maps is 0.96. The SNR turns out to be $\sim 9$ for the same parameters for the foreground subtracted smoothed image, which is slightly smaller than the foreground-free image (the SNR turned out to be $\sim 11$ in that case).

An alternate method of subtracting the foregrounds would be to use a filter in the visibility space such that any frequency-independent component is subtracted out. We have discussed such a filter in Paper I and found that it is quite effective in removing the foregrounds. We implement the same filter in this paper as well and find the results to be similar. The resulting SNR for the fiducial model parameters turns out to be $\sim 10$, which is similar to the other method. We discuss the details of this filtering method in Appendix \ref{filter_fg}.

%%%%%%%%%%%%%%%

\subsection{Realistic maps of the first sources}
\label{res_realistic}

Till now we have been working under the assumption that there is only one source in the FOV. In reality, however, one expects to have multiple sources in the field, and depending on the separation between them there could be significant overlap in the 21-cm patterns. We study these effects using a full cosmological simulation. The steps to generate the realistic maps are briefly described below and one can find the details of the method in \citet{ghara15a, ghara15b}.

\begin{itemize}
\item We first generate the dark matter density and velocity fields at different redshift slices between redshift 20-6 from  a $N-$body simulation using the code {\sc cubep}$^3${\sc m}\footnote{\tt http://wiki.cita.utoronto.ca/mediawiki/index.php/CubePM} \citep{Harnois12} with $2592^3$ particles in a simulation box of size $300 ~h^{-1}$ cMpc. The minimum halo identified using spherical overdensity method is $\sim 4\times 10^9 ~\MSUN $. 

\item We assume that each dark matter halo contains radiating sources. The relation between the stellar mass of the source ($M_{\star}$) and the mass of the hosting halo ($M_{\rm halo}$) is assumed to be,
\begin{equation}
M_{\star}=f_{\star} \left(\frac{{\Omega}_B}{{\Omega}_m} \right) M_{\rm halo},
\label{eq_stell_halo}
\end{equation}
where $f_{\star}$ is the stellar fraction of the baryon in the source. We choose $f_\star = 0.07$ so that the reionization optical depth $\tau=0.0584$ is consistent with the measurement of \citet{2015arXiv150201589P}. In this model, the reionization ends around $z \sim 6.3$.

\item We generate the ionization and temperature maps in the simulation box using a one-dimensional radiative transfer around the sources. We use the pre-generated one-dimensional brightness profiles and a correlation between the temperature and ionization fraction to generate the ionization and temperature maps in the simulation box. The details of the method can be found in \citet{ghara15a}. 

\item We assume that the $\lya$ flux falls as $1/R^2$ with distance $R$. We calculate the  coupling coefficients ($\lya$ coupling, collisional coupling and coupling with the CMBR photons) which are used to generate $\TS$ and $\TB$ maps.

\item We incorporate the effect of the peculiar velocities of the gas in the IGM using cell moving technique \citep{ghara15b}. Finally, we incorporate the light-cone effect, which is described in \citet{ghara15b} in details.

\item  Finally, we re-grid the simulation box to get the desired angular resolution, frequency resolution and observational band width.

\end{itemize}

\begin{figure*}
\begin{center}
\includegraphics[scale=0.55]{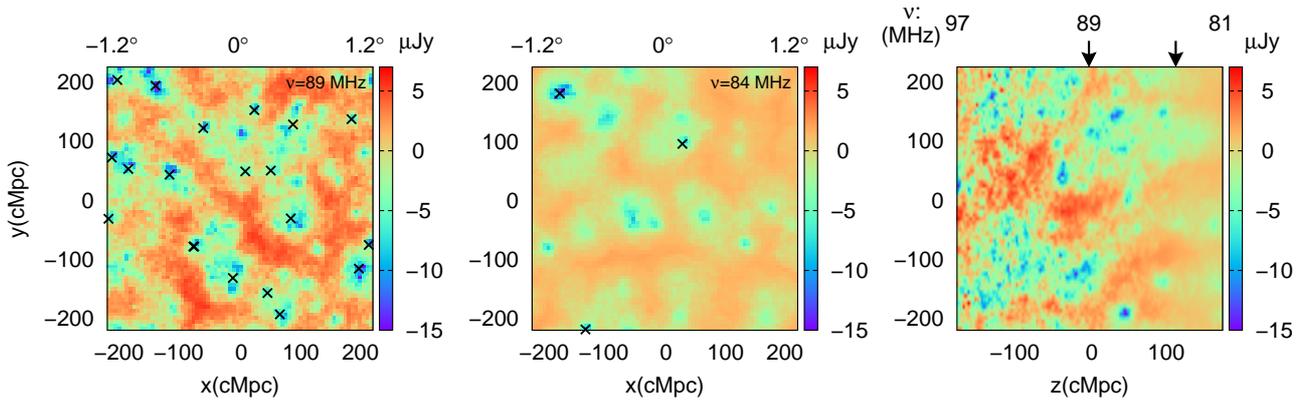}
    \caption{Left-hand panel: $2.4^{\circ}\times 2.4^{\circ}$ map of the 21-cm signal (without smoothing) at the frequency channel $\nu_c=89$ MHz generated from the simulation using the $N-$body simulation and a one-dimensional radiative transfer code. The ``$\times$'' marks represent the angular positions of the sources between a band $\nu_c-0.1$ to $\nu_c+0.1$ MHz.   Middle panel: Same as the left-hand panel, but at the frequency channel corresponding to $\nu=84$ MHz. Right-hand panel: The light-cone map of $\TB$ distribution. The arrows in the top label of the panel show the frequency channels correspond to the maps at the left-hand and the middle panels. The maps include the effect of redshift-space distortion and the light-cone effect. }
   \label{image_p4lcslice}
\end{center}
\end{figure*}

\begin{figure*}
\begin{center}
\includegraphics[scale=0.65]{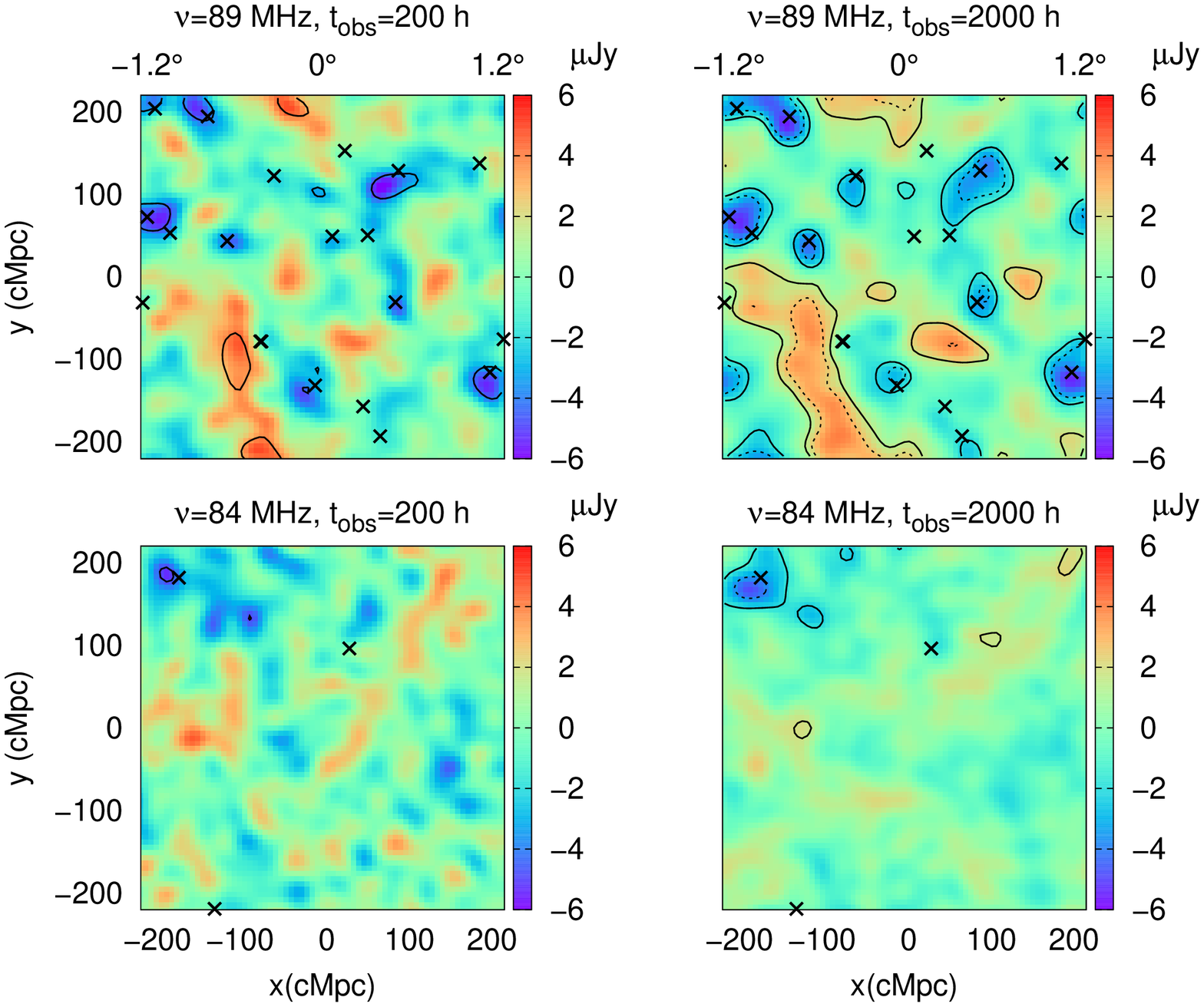}
    \caption{Top left-hand panel: Map of the residual signal and noise after foreground subtraction and smoothing with the Gaussian filter of size 30\arcmin ~at the central frequency channel $\nu_c=89$ MHz. The signal is generated using the method described in section \ref{res_realistic}. The noise corresponds to 200 h of observation, while the smoothing is done with a Gaussian filter of size 30\arcmin.  The ``$\times$'' marks show the angular positions of the sources present within a band $\nu_c-0.1$ to $\nu_c+0.1$ MHz. Top right-hand panel : Same as the top left-hand panel, but for 2000 h of observation. Bottom left-hand panel: Same as the top left-hand panel, but at a different frequency channel which corresponds to $\nu=$ 84 MHz. Bottom right-hand panel: Same as the bottom left-hand panel, but for 2000 h of observation. The solid contours in all the panels correspond to $3-\sigma$ level, while the dotted curves correspond to $5-\sigma$ contours. }
   \label{image_p4contoursim}
\end{center}
\end{figure*}

The left-hand panel of Figure \ref{image_p4lcslice} shows the $2.4^{\circ}\times 2.4^{\circ}$ image of the brightness temperature at the central frequency channel which corresponds to redshift $z=15$. We show the angular positions of the sources by the `$\times$' marks within a band $\nu_c-0.1$ to $\nu_c+0.1$ MHz around the central frequency channel.  One can clearly identify the absorption regions around the sources, however, there is substantial overlap between the individual patterns.  The middle panel of Figure \ref{image_p4lcslice} shows the $\TB$ map at redshift 16. The number of sources drops quite drastically in this frequency channel as we are probing the initial stages of the cosmic dawn. The overlap between the individual patterns too is not that substantial. The decrease in the number of sources towards the lower frequency channels can be seen from the right-hand panel of Figure \ref{image_p4lcslice}, where we show the light-cone from our simulation box.  The signal at the higher frequency end of the box is essentially the strong absorption signal arising from the significant overlap between the $\lya$ photons from the sources. Note that we have subtracted the mean from each frequency channel while making the image and thus, the signal is a combination of emission and absorption regions. As the signal from the channels corresponding to redshifts 15 and 16 are dominated by the absorption regions, the mean of the expected signal is negative.  The regions with positive values of the signal in the left-hand and middle panels of Figure \ref{image_p4lcslice} arise because of subtracting the mean signal from the maps. In reality, these are the $\lya$ deficient regions with almost no 21-cm signal.

The top left-hand  panel of Figure \ref{image_p4contoursim} shows the smoothed image of the residual signal and noise at the central frequency channel after subtracting the foregrounds using the polynomial method. The noise in  the panel corresponds to 200 h of observation and the smoothing is done using the Gaussian filter of size 30\arcmin. One can easily identify that even in the smoothed map, the signal is localized around  the sources. The top right-hand panel of Figure \ref{image_p4contoursim} shows the same but for 2000 h of observation. The signal to noise ratios of these top left-hand and right-hand panels  are 4.8 and 14.2 respectively. The bottom left-hand and right-hand panels are same as the top panels but at a different frequency channel corresponding to $\nu=84$ MHz ($z = 16$). The corresponding SNRs in the bottom panels are 3.3 and 10.1 respectively. The Pearson cross-correlation coefficients $\chi$, in this case, are given in Table \ref{table_chi}. We find that the foreground subtraction method works effectively in this case as well leading to reasonably high values of $\chi$.

We have also shown the $3-\sigma$ and $5-\sigma$ contours of the signal in Figure \ref{image_p4contoursim}. One can see that it is possible to detect the signal at the $3-\sigma$ level within a modest 200 h of observations. The detection can be made more definite in an integration time of 2000 h where the signal is well above the $5-\sigma$ level. This can, in principle, help us in devising strategies for detecting the first sources. For definiteness, let us concentrate on the 84 MHz maps (bottom panels). One can see that, in the given field, one can identify a $3-\sigma$ region near the top-left corner of the map with a 200 h of observations (bottom left-hand panel). Once such a tentative detection of the signal happens in some field, one can attempt longer observation like 2000 h to identify the absorption regions with a larger SNR, as is shown in the bottom right-hand panel. As most of the absorption regions detected in the maps are situated around some sources, one can identify some isolated absorption region on the map to measure the smoothed $\TB$ profile around the corresponding isolated source. Once this smoothed $\TB$ profile is measured with error bars, it can be used to estimate the source parameters using some sophisticated methods like MCMC.

It is possible that the $\lya$ coupling can be quite efficient in the very early stages of galaxy formation, thus reducing the fraction of points that remain $\lya$ deficient at redshifts of observation. In such a ``$\lya$ coupled'' scenario, the 21-cm profile around the sources will be different than what has been considered here \citep{ghara15c}. In addition, the inhomogeneities in  the cosmic density field too can have a significant impact on the signal maps \citep[see e.g.,][]{2000ApJ...528..597T}. In the ``$\lya$ coupled scenario'', a central overdense region followed by an underdense region can provide a $\TB$ profile (after smoothing with a Gaussian filter) similar to that shown in Figure \ref{image_p4tbprof_smt_allpara}. This could lead to an incorrect interpretation of the observations if this degeneracy between the fluctuations in the density field and the $\lya$ radiation is not properly accounted for. A possible way out could be to use targetted follow-up observations using infrared telescopes and determine if there exists a radiation source at a location that is consistent with the 21-cm profile.

\begin{table}
\centering
\begin{tabular}{|c|c|c|}
\hline
Redshift & $t_{\rm obs}$ (h) & $\chi$  \\
\hline
\hline
15 & 200 & 0.85 \\
\hline
15 & 2000 & 0.8 \\
\hline
16 & 200 & 0.89 \\
\hline
16 & 2000 & 0.74 \\
\hline
\end{tabular}
\caption[]{The table shows the Pearson cross-correlation coefficients for different maps from the full cosmological simulation at different redshifts and for different observation time. The coefficient is calculated for the smoothed maps for the signal + noise and the residual signal + noise after foreground subtraction. }
\label{table_chi}
\end{table}

%%%%%%%%%%%%%%%

\section{Summary and discussion}
\label{conc}

We have investigated the detectability of the first sources during the cosmic dawn using imaging techniques through future radio observations with the SKA1-low. Detecting the 21-cm signature of these sources is expected to reveal, at least to some extent, their properties and also the physical state of the surrounding IGM. However, their detection would be significantly challenging because the signal is much too weak compared to the system noise and the astrophysical foregrounds.

Our fiducial source model consists of stars within a galaxy along with a mini-QSO type X-ray source. The model for the sources can be parametrized by several unknown parameters, e.g., the stellar mass ($M_{\star}$), the escape fraction of the UV photons ($f_{\rm esc}$), the ratio of X-ray and UV luminosities ($f_X$), the X-ray spectral index ($\alpha$), the age of the source ($t_{\rm age}$), and the redshift of observation ($z$). In addition, we also need to specify the overdensity of the surrounding IGM ($1+\delta$), assuming it to be uniform. The fiducial values of these parameters are taken to be $M_{\star}=10^7 ~\MSUN, f_{\rm esc}=0.1, f_X=0.05, \alpha =1.5, t_{\rm age}=20$ Myr, $z=15$ and $1 + \delta = 1$. 

We have considered a fiducial observation using the present antenna configuration of the SKA1-low. Assuming that we observe a region at declination $\delta = -30^{\circ}$, we have used the baseline distribution to obtain the ``dirty'' map. We have added the system noise as well as the astrophysical foregrounds (Galactic synchrotron and extragalactic point sources) to the images. Our main aim is to explore whether the images can be used for detecting the signal from the first sources and if one can extract the properties of these sources from the maps.

Our main findings are listed below.

\begin{itemize}

\item If we assume the target source to be isolated, then in the situation where foregrounds can be perfectly subtracted out, it is possible to achieve a signal to noise ratio (SNR) $\sim 11$ for the fiducial source at a redshift of 15 for 2000 h of observation and a frequency resolution of 100 kHz. This SNR is achieved by smoothing the images with a Gaussian filter of size 30\arcmin which helps in reducing the rms of the noise considerably. In general, the SNR increases with increasing width of the Gaussian filter.

\item It is not possible to detect the signal in any reasonable observational time without smoothing the maps. Unfortunately, this smoothing alters the intrinsic brightness temperature profile around the sources which in turn makes it difficult to reliably extract their properties from the maps. We find that it is still possible to constrain the parameters $M_\star$, $f_{\rm esc}$ and $1+\delta$, while it will be difficult to extract any information on $f_X$, $\alpha$ and $t_{\rm age}$ from the smoothed $\TB$ profiles.

\item Although the expected brightness temperature profiles around different types of sources are different, smoothing the maps makes it difficult to distinguish between these sources. In particular, we find that the smoothed profiles of the different X-ray sources, e.g., mini-QSOs and HMXBs, are similar to the case where there are no X-rays from the galaxy.

\item The cosmological 21-cm signal is largely contaminated by the astrophysical foregrounds. In order to account for these, we model the Galactic synchrotron emission and extragalactic point sources \citep{Choudhuri2014MNRAS.445.4351C} and add them to our maps. We then use a third order polynomial fitting method to subtract the foregrounds. We are able to achieve an SNR $\sim 9$ for the fiducial source model which is only $\sim 20\%$ worse than the foreground-free scenario.

\item Since the first galaxies are not expected to form in complete isolation, we generate more realistic signal maps from the output of a $N$-body simulation and using a one-dimensional radiative transfer code \citep{ghara15a}. The reionization history is calibrated to recent Planck measurements of the electron scattering optical depth \citep{2015arXiv150201589P}. We apply the same smoothing and foreground removal technique on these maps as discussed above. The SNR of the map at the redshift 15, after the foregrounds subtraction and smoothing with the fiducial filter,  is $\sim$ 14 (4) for 2000 (200) h of observations. The corresponding SNR value is 10 (3) at redshift 16. This suggests that a possible observation strategy for the SKA1-low could be to observe multiple fields for small observation time like 200 h. If one is able to detect a $3-\sigma$ signal in any of these fields (after smoothing with filters of widths $\sim 30\arcmin$), then one can perform a deeper observation of $\sim 2000$ h and possibly constrain properties of the first sources along with the surrounding IGM.

\end{itemize}

Finally, we discuss some of the aspects of the study which need to be addressed in more details. Although we have modelled the foregrounds in a fairly detailed manner, they can be more complex in the actual case. One probably needs to devise more sophisticated methods to disentangle the signal in that case. Our analysis ignores various other complications, e.g., those arising from instabilities in the ionosphere, calibration of the signal, man-made interference, and instrumental systematics. One possible extension of the present work could be to consider all these complexities and develop a complete pipeline to prepare mock data sets for analysis.

On the modelling aspect, one needs to work out the signal in different reionization scenarios accounting for the uncertainties in the galaxy formation processes at high redshifts. This could include studying the effects of, e.g., the small mass sources of ionization and heating leading to a relatively early overlap of $\lya$ regions \citep{ghara15a}, alternate reionization scenarios driven by quasars \citep{2015ApJ...813L...8M, 2016MNRAS.457.4051K, 2016arXiv160602719M}.

It is also possible to improve the methods used for detecting the signal. In this paper, we have mainly concentrated on the possibility of imaging the 21-cm pattern of the first sources which can be useful, particularly for visual identification, in a situation when the patterns around different sources overlap with each other. However, it is possible that a more efficient search can be performed in the visibility space where the noise is uncorrelated \citep{ghara15c}. In addition, the smoothing filters used in this work have been constructed assuming that we do not have any prior idea of the signal. One could also explore devising more sophisticated filters (e.g., matched filters) which account for the nature of the signal to make a more efficient detection.

\section*{Acknowledgement}
The authors would like to thank Somnath Bharadwaj, Abhik Ghosh, Abhirup Datta, Subhashis Roy, Prasun Dutta and Rohit Sharma for useful discussions and constructive comments on the work. KKD would like to thank DST for support through the project SR/FTP/PS-119/2012 and the University Grant Commission (UGC), India for support through UGC-faculty recharge scheme (UGC-FRP) vide ref. no. F.4-5(137-FRP)/2014(BSR). RG and TRC acknowledge support from the Munich Institute for Astro- and Particle Physics  (MIAPP)  of  the  DFG  cluster of excellence ``Origin and Structure of the Universe''. 

\bibliography{source_image_v4_resubmit}

\begin{thebibliography}{}
\makeatletter
\relax
\def\mn@urlcharsother{\let\do\@makeother \do\$\do\&\do\#\do\^\do\_\do\%\do\~}
\def\mn@doi{\begingroup\mn@urlcharsother \@ifnextchar [ {\mn@doi@}
  {\mn@doi@[]}}
\def\mn@doi@[#1]#2{\def\@tempa{#1}\ifx\@tempa\@empty \href
  {http://dx.doi.org/#2} {doi:#2}\else \href {http://dx.doi.org/#2} {#1}\fi
  \endgroup}
\def\mn@eprint#1#2{\mn@eprint@#1:#2::\@nil}
\def\mn@eprint@arXiv#1{\href {http://arxiv.org/abs/#1} {{\tt arXiv:#1}}}
\def\mn@eprint@dblp#1{\href {http://dblp.uni-trier.de/rec/bibtex/#1.xml}
  {dblp:#1}}
\def\mn@eprint@#1:#2:#3:#4\@nil{\def\@tempa {#1}\def\@tempb {#2}\def\@tempc
  {#3}\ifx \@tempc \@empty \let \@tempc \@tempb \let \@tempb \@tempa \fi \ifx
  \@tempb \@empty \def\@tempb {arXiv}\fi \@ifundefined
  {mn@eprint@\@tempb}{\@tempb:\@tempc}{\expandafter \expandafter \csname
  mn@eprint@\@tempb\endcsname \expandafter{\@tempc}}}

\bibitem[\protect\citeauthoryear{{Ahn}, {Xu}, {Norman}, {Alvarez}  \&
  {Wise}}{{Ahn} et~al.}{2015}]{2015ApJ...802....8A}
{Ahn} K.,  {Xu} H.,  {Norman} M.~L.,  {Alvarez} M.~A.,   {Wise} J.~H.,  2015,
  \mn@doi [\apj] {10.1088/0004-637X/802/1/8}, \href
  {http://adsabs.harvard.edu/abs/2015ApJ...802....8A} {802, 8}

\bibitem[\protect\citeauthoryear{{Ali}, {Bharadwaj}  \& {Chengalur}}{{Ali}
  et~al.}{2008}]{2008MNRAS.385.2166A}
{Ali} S.~S.,  {Bharadwaj} S.,   {Chengalur} J.~N.,  2008, \mn@doi [\mnras]
  {10.1111/j.1365-2966.2008.12984.x}, \href
  {http://adsabs.harvard.edu/abs/2008MNRAS.385.2166A} {385, 2166}

\bibitem[\protect\citeauthoryear{{Alonso}, {Bull}, {Ferreira}  \&
  {Santos}}{{Alonso} et~al.}{2015}]{2015MNRAS.447..400A}
{Alonso} D.,  {Bull} P.,  {Ferreira} P.~G.,   {Santos} M.~G.,  2015, \mn@doi
  [\mnras] {10.1093/mnras/stu2474}, \href
  {http://adsabs.harvard.edu/abs/2015MNRAS.447..400A} {447, 400}

\bibitem[\protect\citeauthoryear{{Alvarez}, {Wise}  \& {Abel}}{{Alvarez}
  et~al.}{2009}]{2009ApJ...701L.133A}
{Alvarez} M.~A.,  {Wise} J.~H.,   {Abel} T.,  2009, \mn@doi [\apjl]
  {10.1088/0004-637X/701/2/L133}, \href
  {http://adsabs.harvard.edu/abs/2009ApJ...701L.133A} {701, L133}

\bibitem[\protect\citeauthoryear{{Alvarez}, {Pen}  \& {Chang}}{{Alvarez}
  et~al.}{2010}]{Alvarez10}
{Alvarez} M.~A.,  {Pen} U.-L.,   {Chang} T.-C.,  2010, \mn@doi [\apjl]
  {10.1088/2041-8205/723/1/L17}, \href
  {http://adsabs.harvard.edu/abs/2010ApJ...723L..17A} {723, L17}

\bibitem[\protect\citeauthoryear{{Ansari} et~al.,}{{Ansari}
  et~al.}{2012}]{2012A&A...540A.129A}
{Ansari} R.,  et~al., 2012, \mn@doi [\aap] {10.1051/0004-6361/201117837}, \href
  {http://adsabs.harvard.edu/abs/2012A%26A...540A.129A} {540, A129}

\bibitem[\protect\citeauthoryear{{Baek}, {Di Matteo}, {Semelin}, {Combes}  \&
  {Revaz}}{{Baek} et~al.}{2009}]{baek09}
{Baek} S.,  {Di Matteo} P.,  {Semelin} B.,  {Combes} F.,   {Revaz} Y.,  2009,
  \mn@doi [\aap] {10.1051/0004-6361:200810757}, \href
  {http://adsabs.harvard.edu/abs/2009A%26A...495..389B} {495, 389}

\bibitem[\protect\citeauthoryear{{Bouwens} et~al.,}{{Bouwens}
  et~al.}{2015}]{Bouwens15}
{Bouwens} R.~J.,  et~al., 2015, \mn@doi [\apj] {10.1088/0004-637X/803/1/34},
  \href {http://adsabs.harvard.edu/abs/2015ApJ...803...34B} {803, 34}

\bibitem[\protect\citeauthoryear{{Bowman} et~al.,}{{Bowman}
  et~al.}{2013}]{bowman13}
{Bowman} J.~D.,  et~al., 2013, \mn@doi [\pasa] {10.1017/pas.2013.009}, \href
  {http://adsabs.harvard.edu/abs/2013PASA...30...31B} {30, e031}

\bibitem[\protect\citeauthoryear{{Bromm} \& {Loeb}}{{Bromm} \&
  {Loeb}}{2003}]{2003ApJ...596...34B}
{Bromm} V.,  {Loeb} A.,  2003, \mn@doi [\apj] {10.1086/377529}, \href
  {http://adsabs.harvard.edu/abs/2003ApJ...596...34B} {596, 34}

\bibitem[\protect\citeauthoryear{{Cen}}{{Cen}}{2006}]{2006ApJ...648...47C}
{Cen} R.,  2006, \mn@doi [\apj] {10.1086/505631}, \href
  {http://adsabs.harvard.edu/abs/2006ApJ...648...47C} {648, 47}

\bibitem[\protect\citeauthoryear{{Chapman} et~al.,}{{Chapman}
  et~al.}{2013}]{2013MNRAS.429..165C}
{Chapman} E.,  et~al., 2013, \mn@doi [\mnras] {10.1093/mnras/sts333}, \href
  {http://adsabs.harvard.edu/abs/2013MNRAS.429..165C} {429, 165}

\bibitem[\protect\citeauthoryear{{Chen} \& {Miralda-Escud{\'e}}}{{Chen} \&
  {Miralda-Escud{\'e}}}{2008}]{2008ApJ...684...18C}
{Chen} X.,  {Miralda-Escud{\'e}} J.,  2008, \mn@doi [\apj] {10.1086/528941},
  \href {http://adsabs.harvard.edu/abs/2008ApJ...684...18C} {684, 18}

\bibitem[\protect\citeauthoryear{{Choudhuri}, {Bharadwaj}, {Ghosh}  \&
  {Ali}}{{Choudhuri} et~al.}{2014}]{Choudhuri2014MNRAS.445.4351C}
{Choudhuri} S.,  {Bharadwaj} S.,  {Ghosh} A.,   {Ali} S.~S.,  2014, \mn@doi
  [\mnras] {10.1093/mnras/stu2027}, \href
  {http://adsabs.harvard.edu/abs/2014MNRAS.445.4351C} {445, 4351}

\bibitem[\protect\citeauthoryear{{Choudhury}, {Haehnelt}  \&
  {Regan}}{{Choudhury} et~al.}{2009}]{choudhury09}
{Choudhury} T.~R.,  {Haehnelt} M.~G.,   {Regan} J.,  2009, \mn@doi [\mnras]
  {10.1111/j.1365-2966.2008.14383.x}, \href
  {http://adsabs.harvard.edu/abs/2009MNRAS.394..960C} {394, 960}

\bibitem[\protect\citeauthoryear{{Chuzhoy}, {Alvarez}  \& {Shapiro}}{{Chuzhoy}
  et~al.}{2006}]{2006ApJ...648L...1C}
{Chuzhoy} L.,  {Alvarez} M.~A.,   {Shapiro} P.~R.,  2006, \mn@doi [\apjl]
  {10.1086/507626}, \href {http://adsabs.harvard.edu/abs/2006ApJ...648L...1C}
  {648, L1}

\bibitem[\protect\citeauthoryear{{Datta}, {Bharadwaj}  \& {Choudhury}}{{Datta}
  et~al.}{2007}]{kanan2007MNRAS.382..809D}
{Datta} K.~K.,  {Bharadwaj} S.,   {Choudhury} T.~R.,  2007, \mn@doi [\mnras]
  {10.1111/j.1365-2966.2007.12421.x}, \href
  {http://adsabs.harvard.edu/abs/2007MNRAS.382..809D} {382, 809}

\bibitem[\protect\citeauthoryear{{Datta}, {Majumdar}, {Bharadwaj}  \&
  {Choudhury}}{{Datta} et~al.}{2008}]{2008MNRAS.391.1900D}
{Datta} K.~K.,  {Majumdar} S.,  {Bharadwaj} S.,   {Choudhury} T.~R.,  2008,
  \mn@doi [\mnras] {10.1111/j.1365-2966.2008.14008.x}, \href
  {http://adsabs.harvard.edu/abs/2008MNRAS.391.1900D} {391, 1900}

\bibitem[\protect\citeauthoryear{{Datta}, {Bharadwaj}  \& {Choudhury}}{{Datta}
  et~al.}{2009}]{2009MNRAS.399L.132D}
{Datta} K.~K.,  {Bharadwaj} S.,   {Choudhury} T.~R.,  2009, \mn@doi [\mnras]
  {10.1111/j.1745-3933.2009.00739.x}, \href
  {http://adsabs.harvard.edu/abs/2009MNRAS.399L.132D} {399, L132}

\bibitem[\protect\citeauthoryear{{Datta}, {Friedrich}, {Mellema}, {Iliev}  \&
  {Shapiro}}{{Datta} et~al.}{2012a}]{datta2012a}
{Datta} K.~K.,  {Friedrich} M.~M.,  {Mellema} G.,  {Iliev} I.~T.,   {Shapiro}
  P.~R.,  2012a, \mn@doi [\mnras] {10.1111/j.1365-2966.2012.21268.x}, \href
  {http://adsabs.harvard.edu/abs/2012MNRAS.424..762D} {424, 762}

\bibitem[\protect\citeauthoryear{{Datta}, {Mellema}, {Mao}, {Iliev}, {Shapiro}
  \& {Ahn}}{{Datta} et~al.}{2012b}]{Datta2012b}
{Datta} K.~K.,  {Mellema} G.,  {Mao} Y.,  {Iliev} I.~T.,  {Shapiro} P.~R.,
  {Ahn} K.,  2012b, \mn@doi [\mnras] {10.1111/j.1365-2966.2012.21293.x}, \href
  {http://adsabs.harvard.edu/abs/2012MNRAS.424.1877D} {424, 1877}

\bibitem[\protect\citeauthoryear{{Di Matteo}, {Perna}, {Abel}  \& {Rees}}{{Di
  Matteo} et~al.}{2002}]{2002ApJ...564..576D}
{Di Matteo} T.,  {Perna} R.,  {Abel} T.,   {Rees} M.~J.,  2002, \mn@doi [\apj]
  {10.1086/324293}, \href {http://cdsads.u-strasbg.fr/abs/2002ApJ...564..576D}
  {564, 576}

\bibitem[\protect\citeauthoryear{{Ellis} et~al.,}{{Ellis}
  et~al.}{2013}]{Ellis13}
{Ellis} R.~S.,  et~al., 2013, \mn@doi [\apjl] {10.1088/2041-8205/763/1/L7},
  \href {http://adsabs.harvard.edu/abs/2013ApJ...763L...7E} {763, L7}

\bibitem[\protect\citeauthoryear{{Fan} et~al.,}{{Fan} et~al.}{2006}]{Fan06b}
{Fan} X.,  et~al., 2006, \mn@doi [\aj] {10.1086/504836}, \href
  {http://adsabs.harvard.edu/abs/2006AJ....132..117F} {132, 117}

\bibitem[\protect\citeauthoryear{{Fialkov}, {Barkana}  \& {Visbal}}{{Fialkov}
  et~al.}{2014}]{Fialkov14}
{Fialkov} A.,  {Barkana} R.,   {Visbal} E.,  2014, \mn@doi [\nat]
  {10.1038/nature12999}, \href
  {http://adsabs.harvard.edu/abs/2014Natur.506..197F} {506, 197}

\bibitem[\protect\citeauthoryear{{Finkelstein}, {Rhoads}, {Malhotra}  \&
  {Grogin}}{{Finkelstein} et~al.}{2009}]{Finkelstein09b}
{Finkelstein} S.~L.,  {Rhoads} J.~E.,  {Malhotra} S.,   {Grogin} N.,  2009,
  \mn@doi [\apj] {10.1088/0004-637X/691/1/465}, \href
  {http://adsabs.harvard.edu/abs/2009ApJ...691..465F} {691, 465}

\bibitem[\protect\citeauthoryear{{Fioc} \& {Rocca-Volmerange}}{{Fioc} \&
  {Rocca-Volmerange}}{1997}]{Fioc97}
{Fioc} M.,  {Rocca-Volmerange} B.,  1997, \aap, \href
  {http://adsabs.harvard.edu/abs/1997A%26A...326..950F} {326, 950}

\bibitem[\protect\citeauthoryear{{Fragos} et~al.,}{{Fragos}
  et~al.}{2013a}]{frag1}
{Fragos} T.,  et~al., 2013a, \mn@doi [\apj] {10.1088/0004-637X/764/1/41}, \href
  {http://adsabs.harvard.edu/abs/2013ApJ...764...41F} {764, 41}

\bibitem[\protect\citeauthoryear{{Fragos}, {Lehmer}, {Naoz}, {Zezas}  \&
  {Basu-Zych}}{{Fragos} et~al.}{2013b}]{frag2}
{Fragos} T.,  {Lehmer} B.~D.,  {Naoz} S.,  {Zezas} A.,   {Basu-Zych} A.,
  2013b, \mn@doi [\apjl] {10.1088/2041-8205/776/2/L31}, \href
  {http://adsabs.harvard.edu/abs/2013ApJ...776L..31F} {776, L31}

\bibitem[\protect\citeauthoryear{{Fukugita} \& {Kawasaki}}{{Fukugita} \&
  {Kawasaki}}{1994}]{1994MNRAS.269..563F}
{Fukugita} M.,  {Kawasaki} M.,  1994, \mn@doi [\mnras]
  {10.1093/mnras/269.3.563}, \href
  {http://adsabs.harvard.edu/abs/1994MNRAS.269..563F} {269, 563}

\bibitem[\protect\citeauthoryear{{Furlanetto}, {Zaldarriaga}  \&
  {Hernquist}}{{Furlanetto} et~al.}{2004}]{furlanetto04}
{Furlanetto} S.~R.,  {Zaldarriaga} M.,   {Hernquist} L.,  2004, \mn@doi [\apj]
  {10.1086/423028}, \href {http://adsabs.harvard.edu/abs/2004ApJ...613...16F}
  {613, 16}

\bibitem[\protect\citeauthoryear{{Geil} \& {Wyithe}}{{Geil} \&
  {Wyithe}}{2008}]{2008MNRAS.386.1683G}
{Geil} P.~M.,  {Wyithe} J.~S.~B.,  2008, \mn@doi [\mnras]
  {10.1111/j.1365-2966.2008.13159.x}, \href
  {http://adsabs.harvard.edu/abs/2008MNRAS.386.1683G} {386, 1683}

\bibitem[\protect\citeauthoryear{{Ghara}, {Choudhury}  \& {Datta}}{{Ghara}
  et~al.}{2015a}]{ghara15a}
{Ghara} R.,  {Choudhury} T.~R.,   {Datta} K.~K.,  2015a, \mn@doi [\mnras]
  {10.1093/mnras/stu2512}, \href
  {http://adsabs.harvard.edu/abs/2015MNRAS.447.1806G} {447, 1806}

\bibitem[\protect\citeauthoryear{{Ghara}, {Datta}  \& {Choudhury}}{{Ghara}
  et~al.}{2015b}]{ghara15b}
{Ghara} R.,  {Datta} K.~K.,   {Choudhury} T.~R.,  2015b, \mn@doi [\mnras]
  {10.1093/mnras/stv1855}, \href
  {http://adsabs.harvard.edu/abs/2015MNRAS.453.3143G} {453, 3143}

\bibitem[\protect\citeauthoryear{{Ghara}, {Choudhury}  \& {Datta}}{{Ghara}
  et~al.}{2016}]{ghara15c}
{Ghara} R.,  {Choudhury} T.~R.,   {Datta} K.~K.,  2016, \mn@doi [\mnras]
  {10.1093/mnras/stw953}, \href
  {http://adsabs.harvard.edu/abs/2016MNRAS.460..827G} {460, 827}

\bibitem[\protect\citeauthoryear{{Ghosh}, {Bharadwaj}, {Ali}  \&
  {Chengalur}}{{Ghosh} et~al.}{2011}]{2011MNRAS.418.2584G}
{Ghosh} A.,  {Bharadwaj} S.,  {Ali} S.~S.,   {Chengalur} J.~N.,  2011, \mn@doi
  [\mnras] {10.1111/j.1365-2966.2011.19649.x}, \href
  {http://adsabs.harvard.edu/abs/2011MNRAS.418.2584G} {418, 2584}

\bibitem[\protect\citeauthoryear{{Ghosh}, {Prasad}, {Bharadwaj}, {Ali}  \&
  {Chengalur}}{{Ghosh} et~al.}{2012}]{ghosh12}
{Ghosh} A.,  {Prasad} J.,  {Bharadwaj} S.,  {Ali} S.~S.,   {Chengalur} J.~N.,
  2012, \mn@doi [\mnras] {10.1111/j.1365-2966.2012.21889.x}, \href
  {http://adsabs.harvard.edu/abs/2012MNRAS.426.3295G} {426, 3295}

\bibitem[\protect\citeauthoryear{{Gleser}, {Nusser}  \& {Benson}}{{Gleser}
  et~al.}{2008}]{2008MNRAS.391..383G}
{Gleser} L.,  {Nusser} A.,   {Benson} A.~J.,  2008, \mn@doi [\mnras]
  {10.1111/j.1365-2966.2008.13897.x}, \href
  {http://adsabs.harvard.edu/abs/2008MNRAS.391..383G} {391, 383}

\bibitem[\protect\citeauthoryear{{Greif}, {Glover}, {Bromm}  \&
  {Klessen}}{{Greif} et~al.}{2010}]{2010ApJ...716..510G}
{Greif} T.~H.,  {Glover} S.~C.~O.,  {Bromm} V.,   {Klessen} R.~S.,  2010,
  \mn@doi [\apj] {10.1088/0004-637X/716/1/510}, \href
  {http://adsabs.harvard.edu/abs/2010ApJ...716..510G} {716, 510}

\bibitem[\protect\citeauthoryear{{Gu}, {Xu}, {Wang}, {An}  \& {Chen}}{{Gu}
  et~al.}{2013}]{2013ApJ...773...38G}
{Gu} J.,  {Xu} H.,  {Wang} J.,  {An} T.,   {Chen} W.,  2013, \mn@doi [\apj]
  {10.1088/0004-637X/773/1/38}, \href
  {http://adsabs.harvard.edu/abs/2013ApJ...773...38G} {773, 38}

\bibitem[\protect\citeauthoryear{{Harker} et~al.,}{{Harker}
  et~al.}{2010}]{2010MNRAS.405.2492H}
{Harker} G.,  et~al., 2010, \mn@doi [\mnras]
  {10.1111/j.1365-2966.2010.16628.x}, \href
  {http://adsabs.harvard.edu/abs/2010MNRAS.405.2492H} {405, 2492}

\bibitem[\protect\citeauthoryear{{Harnois-D{\'e}raps}, {Pen}, {Iliev}, {Merz},
  {Emberson}  \& {Desjacques}}{{Harnois-D{\'e}raps} et~al.}{2013}]{Harnois12}
{Harnois-D{\'e}raps} J.,  {Pen} U.-L.,  {Iliev} I.~T.,  {Merz} H.,  {Emberson}
  J.~D.,   {Desjacques} V.,  2013, \mn@doi [\mnras] {10.1093/mnras/stt1591},
  \href {http://adsabs.harvard.edu/abs/2013MNRAS.436..540H} {436, 540}

\bibitem[\protect\citeauthoryear{{Higgins} \& {Meiksin}}{{Higgins} \&
  {Meiksin}}{2012}]{2012MNRAS.426.2380H}
{Higgins} J.,  {Meiksin} A.,  2012, \mn@doi [\mnras]
  {10.1111/j.1365-2966.2012.21917.x}, \href
  {http://adsabs.harvard.edu/abs/2012MNRAS.426.2380H} {426, 2380}

\bibitem[\protect\citeauthoryear{{Hu}, {Cowie}, {Barger}, {Capak}, {Kakazu}  \&
  {Trouille}}{{Hu} et~al.}{2010}]{Hu10}
{Hu} E.~M.,  {Cowie} L.~L.,  {Barger} A.~J.,  {Capak} P.,  {Kakazu} Y.,
  {Trouille} L.,  2010, \mn@doi [\apj] {10.1088/0004-637X/725/1/394}, \href
  {http://adsabs.harvard.edu/abs/2010ApJ...725..394H} {725, 394}

\bibitem[\protect\citeauthoryear{{Iliev} et~al.,}{{Iliev}
  et~al.}{2006}]{Iliev2006}
{Iliev} I.~T.,  et~al., 2006, \mn@doi [\mnras]
  {10.1111/j.1365-2966.2006.10775.x}, \href
  {http://adsabs.harvard.edu/abs/2006MNRAS.371.1057I} {371, 1057}

\bibitem[\protect\citeauthoryear{{Jeli{\'c}} et~al.,}{{Jeli{\'c}}
  et~al.}{2008}]{jelic08}
{Jeli{\'c}} V.,  et~al., 2008, \mn@doi [\mnras]
  {10.1111/j.1365-2966.2008.13634.x}, \href
  {http://adsabs.harvard.edu/abs/2008MNRAS.389.1319J} {389, 1319}

\bibitem[\protect\citeauthoryear{{Kaaret}}{{Kaaret}}{2014}]{2014MNRAS.440L..26K}
{Kaaret} P.,  2014, \mn@doi [\mnras] {10.1093/mnrasl/slu018}, \href
  {http://adsabs.harvard.edu/abs/2014MNRAS.440L..26K} {440, L26}

\bibitem[\protect\citeauthoryear{{Kashikawa} et~al.,}{{Kashikawa}
  et~al.}{2011}]{Kashikawa11}
{Kashikawa} N.,  et~al., 2011, \mn@doi [\apj] {10.1088/0004-637X/734/2/119},
  \href {http://adsabs.harvard.edu/abs/2011ApJ...734..119K} {734, 119}

\bibitem[\protect\citeauthoryear{{Khaire}, {Srianand}, {Choudhury}  \&
  {Gaikwad}}{{Khaire} et~al.}{2016}]{2016MNRAS.457.4051K}
{Khaire} V.,  {Srianand} R.,  {Choudhury} T.~R.,   {Gaikwad} P.,  2016, \mn@doi
  [\mnras] {10.1093/mnras/stw192}, \href
  {http://adsabs.harvard.edu/abs/2016MNRAS.457.4051K} {457, 4051}

\bibitem[\protect\citeauthoryear{{Knevitt}, {Wynn}, {Power}  \&
  {Bolton}}{{Knevitt} et~al.}{2014}]{2014MNRAS.445.2034K}
{Knevitt} G.,  {Wynn} G.~A.,  {Power} C.,   {Bolton} J.~S.,  2014, \mn@doi
  [\mnras] {10.1093/mnras/stu1803}, \href
  {http://adsabs.harvard.edu/abs/2014MNRAS.445.2034K} {445, 2034}

\bibitem[\protect\citeauthoryear{{Lai}, {Huang}, {Fazio}, {Cowie}, {Hu}  \&
  {Kakazu}}{{Lai} et~al.}{2007}]{Lai07}
{Lai} K.,  {Huang} J.-S.,  {Fazio} G.,  {Cowie} L.~L.,  {Hu} E.~M.,   {Kakazu}
  Y.,  2007, \mn@doi [\apj] {10.1086/510285}, \href
  {http://adsabs.harvard.edu/abs/2007ApJ...655..704L} {655, 704}

\bibitem[\protect\citeauthoryear{{Laor}, {Fiore}, {Elvis}, {Wilkes}  \&
  {McDowell}}{{Laor} et~al.}{1997}]{laor97}
{Laor} A.,  {Fiore} F.,  {Elvis} M.,  {Wilkes} B.~J.,   {McDowell} J.~C.,
  1997, \apj, \href {http://adsabs.harvard.edu/abs/1997ApJ...477...93L} {477,
  93}

\bibitem[\protect\citeauthoryear{{Liu}, {Tegmark}, {Bowman}, {Hewitt}  \&
  {Zaldarriaga}}{{Liu} et~al.}{2009}]{2009MNRAS.398..401L}
{Liu} A.,  {Tegmark} M.,  {Bowman} J.,  {Hewitt} J.,   {Zaldarriaga} M.,  2009,
  \mn@doi [\mnras] {10.1111/j.1365-2966.2009.15156.x}, \href
  {http://adsabs.harvard.edu/abs/2009MNRAS.398..401L} {398, 401}

\bibitem[\protect\citeauthoryear{{Madau} \& {Haardt}}{{Madau} \&
  {Haardt}}{2015}]{2015ApJ...813L...8M}
{Madau} P.,  {Haardt} F.,  2015, \mn@doi [\apjl] {10.1088/2041-8205/813/1/L8},
  \href {http://adsabs.harvard.edu/abs/2015ApJ...813L...8M} {813, L8}

\bibitem[\protect\citeauthoryear{{Maio}, {Ciardi}, {Dolag}, {Tornatore}  \&
  {Khochfar}}{{Maio} et~al.}{2010}]{2010MNRAS.407.1003M}
{Maio} U.,  {Ciardi} B.,  {Dolag} K.,  {Tornatore} L.,   {Khochfar} S.,  2010,
  \mn@doi [\mnras] {10.1111/j.1365-2966.2010.17003.x}, \href
  {http://adsabs.harvard.edu/abs/2010MNRAS.407.1003M} {407, 1003}

\bibitem[\protect\citeauthoryear{{Majumdar}, {Bharadwaj}, {Datta}  \&
  {Choudhury}}{{Majumdar} et~al.}{2011}]{2011MNRAS.413.1409M}
{Majumdar} S.,  {Bharadwaj} S.,  {Datta} K.~K.,   {Choudhury} T.~R.,  2011,
  \mn@doi [\mnras] {10.1111/j.1365-2966.2011.18223.x}, \href
  {http://adsabs.harvard.edu/abs/2011MNRAS.413.1409M} {413, 1409}

\bibitem[\protect\citeauthoryear{{Majumdar}, {Bharadwaj}  \&
  {Choudhury}}{{Majumdar} et~al.}{2012}]{2012MNRAS.426.3178M}
{Majumdar} S.,  {Bharadwaj} S.,   {Choudhury} T.~R.,  2012, \mn@doi [\mnras]
  {10.1111/j.1365-2966.2012.21914.x}, \href
  {http://adsabs.harvard.edu/abs/2012MNRAS.426.3178M} {426, 3178}

\bibitem[\protect\citeauthoryear{{McQuinn}, {Zahn}, {Zaldarriaga}, {Hernquist}
  \& {Furlanetto}}{{McQuinn} et~al.}{2006}]{2006ApJ...653..815M}
{McQuinn} M.,  {Zahn} O.,  {Zaldarriaga} M.,  {Hernquist} L.,   {Furlanetto}
  S.~R.,  2006, \mn@doi [\apj] {10.1086/505167}, \href
  {http://adsabs.harvard.edu/abs/2006ApJ...653..815M} {653, 815}

\bibitem[\protect\citeauthoryear{{McQuinn}, {Lidz}, {Zahn}, {Dutta},
  {Hernquist}  \& {Zaldarriaga}}{{McQuinn} et~al.}{2007}]{McQuinn2007}
{McQuinn} M.,  {Lidz} A.,  {Zahn} O.,  {Dutta} S.,  {Hernquist} L.,
  {Zaldarriaga} M.,  2007, \mn@doi [\mnras] {10.1111/j.1365-2966.2007.11489.x},
  \href {http://adsabs.harvard.edu/abs/2007MNRAS.377.1043M} {377, 1043}

\bibitem[\protect\citeauthoryear{{Mellema}, {Iliev}, {Pen}  \&
  {Shapiro}}{{Mellema} et~al.}{2006}]{mellema06}
{Mellema} G.,  {Iliev} I.~T.,  {Pen} U.-L.,   {Shapiro} P.~R.,  2006, \mn@doi
  [\mnras] {10.1111/j.1365-2966.2006.10919.x}, \href
  {http://adsabs.harvard.edu/abs/2006MNRAS.372..679M} {372, 679}

\bibitem[\protect\citeauthoryear{{Mellema}, {Koopmans}, {Shukla}, {Datta},
  {Mesinger}  \& {Majumdar}}{{Mellema} et~al.}{2015}]{2015aska.confE..10M}
{Mellema} G.,  {Koopmans} L.,  {Shukla} H.,  {Datta} K.~K.,  {Mesinger} A.,
  {Majumdar} S.,  2015, Advancing Astrophysics with the Square Kilometre Array
  (AASKA14), \href {http://adsabs.harvard.edu/abs/2015aska.confE..10M} {p.~10}

\bibitem[\protect\citeauthoryear{{Mesinger} \& {Furlanetto}}{{Mesinger} \&
  {Furlanetto}}{2007}]{mesinger07}
{Mesinger} A.,  {Furlanetto} S.,  2007, \mn@doi [\apj] {10.1086/521806}, \href
  {http://adsabs.harvard.edu/abs/2007ApJ...669..663M} {669, 663}

\bibitem[\protect\citeauthoryear{{Meynet} \& {Maeder}}{{Meynet} \&
  {Maeder}}{2005}]{Meyn05}
{Meynet} G.,  {Maeder} A.,  2005, \mn@doi [\aap] {10.1051/0004-6361:20047106},
  \href {http://adsabs.harvard.edu/abs/2005A%26A...429..581M} {429, 581}

\bibitem[\protect\citeauthoryear{{Mirabel}, {Dijkstra}, {Laurent}, {Loeb}  \&
  {Pritchard}}{{Mirabel} et~al.}{2011}]{2011A&A...528A.149M}
{Mirabel} I.~F.,  {Dijkstra} M.,  {Laurent} P.,  {Loeb} A.,   {Pritchard}
  J.~R.,  2011, \mn@doi [\aap] {10.1051/0004-6361/201016357}, \href
  {http://adsabs.harvard.edu/abs/2011A%26A...528A.149M} {528, A149}

\bibitem[\protect\citeauthoryear{{Mitra}, {Choudhury}  \& {Ferrara}}{{Mitra}
  et~al.}{2016}]{2016arXiv160602719M}
{Mitra} S.,  {Choudhury} T.~R.,   {Ferrara} A.,  2016, preprint, \href
  {http://adsabs.harvard.edu/abs/2016arXiv160602719M} {} (\mn@eprint {arXiv}
  {1606.02719})

\bibitem[\protect\citeauthoryear{{O'Shea}, {Wise}, {Xu}  \& {Norman}}{{O'Shea}
  et~al.}{2015}]{2015ApJ...807L..12O}
{O'Shea} B.~W.,  {Wise} J.~H.,  {Xu} H.,   {Norman} M.~L.,  2015, \mn@doi
  [\apjl] {10.1088/2041-8205/807/1/L12}, \href
  {http://adsabs.harvard.edu/abs/2015ApJ...807L..12O} {807, L12}

\bibitem[\protect\citeauthoryear{{Ouchi} et~al.,}{{Ouchi}
  et~al.}{2010}]{Ouchi10}
{Ouchi} M.,  et~al., 2010, \mn@doi [\apj] {10.1088/0004-637X/723/1/869}, \href
  {http://adsabs.harvard.edu/abs/2010ApJ...723..869O} {723, 869}

\bibitem[\protect\citeauthoryear{{Paciga} et~al.,}{{Paciga}
  et~al.}{2013}]{paciga13}
{Paciga} G.,  et~al., 2013, \mn@doi [\mnras] {10.1093/mnras/stt753}, \href
  {http://adsabs.harvard.edu/abs/2013MNRAS.433..639P} {433, 639}

\bibitem[\protect\citeauthoryear{{Paranjape} \& {Choudhury}}{{Paranjape} \&
  {Choudhury}}{2014}]{2014MNRAS.442.1470P}
{Paranjape} A.,  {Choudhury} T.~R.,  2014, \mn@doi [\mnras]
  {10.1093/mnras/stu911}, \href
  {http://adsabs.harvard.edu/abs/2014MNRAS.442.1470P} {442, 1470}

\bibitem[\protect\citeauthoryear{{Parsons} et~al.,}{{Parsons}
  et~al.}{2014}]{parsons13}
{Parsons} A.~R.,  et~al., 2014, \mn@doi [\apj] {10.1088/0004-637X/788/2/106},
  \href {http://adsabs.harvard.edu/abs/2014ApJ...788..106P} {788, 106}

\bibitem[\protect\citeauthoryear{{Pawlik}, {Milosavljevi{\'c}}  \&
  {Bromm}}{{Pawlik} et~al.}{2011}]{2011ApJ...731...54P}
{Pawlik} A.~H.,  {Milosavljevi{\'c}} M.,   {Bromm} V.,  2011, \mn@doi [\apj]
  {10.1088/0004-637X/731/1/54}, \href
  {http://adsabs.harvard.edu/abs/2011ApJ...731...54P} {731, 54}

\bibitem[\protect\citeauthoryear{{Petrovic} \& {Oh}}{{Petrovic} \&
  {Oh}}{2011}]{2011MNRAS.413.2103P}
{Petrovic} N.,  {Oh} S.~P.,  2011, \mn@doi [\mnras]
  {10.1111/j.1365-2966.2011.18276.x}, \href
  {http://adsabs.harvard.edu/abs/2011MNRAS.413.2103P} {413, 2103}

\bibitem[\protect\citeauthoryear{{Planck Collaboration} et~al.,}{{Planck
  Collaboration} et~al.}{2014}]{Planck2013}
{Planck Collaboration} et~al., 2014, \mn@doi [\aap]
  {10.1051/0004-6361/201321591}, \href
  {http://adsabs.harvard.edu/abs/2014A%26A...571A..16P} {571, A16}

\bibitem[\protect\citeauthoryear{{Planck Collaboration} et~al.,}{{Planck
  Collaboration} et~al.}{2015}]{2015arXiv150201589P}
{Planck Collaboration} et~al., 2015, preprint, \href
  {http://adsabs.harvard.edu/abs/2015arXiv150201589P} {} (\mn@eprint {arXiv}
  {1502.01589})

\bibitem[\protect\citeauthoryear{{Platania}, {Bensadoun}, {Bersanelli}, {De
  Amici}, {Kogut}, {Levin}, {Maino}  \& {Smoot}}{{Platania}
  et~al.}{1998}]{1998ApJ...505..473P}
{Platania} P.,  {Bensadoun} M.,  {Bersanelli} M.,  {De Amici} G.,  {Kogut} A.,
  {Levin} S.,  {Maino} D.,   {Smoot} G.~F.,  1998, \mn@doi [\apj]
  {10.1086/306175}, \href {http://adsabs.harvard.edu/abs/1998ApJ...505..473P}
  {505, 473}

\bibitem[\protect\citeauthoryear{{Rix} et~al.,}{{Rix} et~al.}{2004}]{Rix04}
{Rix} H.-W.,  et~al., 2004, \mn@doi [\apjs] {10.1086/420885}, \href
  {http://ads.ari.uni-heidelberg.de/abs/2004ApJS..152..163R} {152, 163}

\bibitem[\protect\citeauthoryear{{Santos}, {Cooray}  \& {Knox}}{{Santos}
  et~al.}{2005}]{2005ApJ...625..575S}
{Santos} M.~G.,  {Cooray} A.,   {Knox} L.,  2005, \mn@doi [\apj]
  {10.1086/429857}, \href {http://adsabs.harvard.edu/abs/2005ApJ...625..575S}
  {625, 575}

\bibitem[\protect\citeauthoryear{{Santos}, {Amblard}, {Pritchard}, {Trac},
  {Cen}  \& {Cooray}}{{Santos} et~al.}{2008}]{santos08}
{Santos} M.~G.,  {Amblard} A.,  {Pritchard} J.,  {Trac} H.,  {Cen} R.,
  {Cooray} A.,  2008, \mn@doi [\apj] {10.1086/592487}, \href
  {http://adsabs.harvard.edu/abs/2008ApJ...689....1S} {689, 1}

\bibitem[\protect\citeauthoryear{{Semelin}, {Combes}  \& {Baek}}{{Semelin}
  et~al.}{2007}]{2007A&A...474..365S}
{Semelin} B.,  {Combes} F.,   {Baek} S.,  2007, \mn@doi [\aap]
  {10.1051/0004-6361:20077965}, \href
  {http://adsabs.harvard.edu/abs/2007A%26A...474..365S} {474, 365}

\bibitem[\protect\citeauthoryear{{Shin}, {Trac}  \& {Cen}}{{Shin}
  et~al.}{2008}]{shin2008}
{Shin} M.-S.,  {Trac} H.,   {Cen} R.,  2008, \mn@doi [\apj] {10.1086/588247},
  \href {http://adsabs.harvard.edu/abs/2008ApJ...681..756S} {681, 756}

\bibitem[\protect\citeauthoryear{{Stacy}, {Greif}  \& {Bromm}}{{Stacy}
  et~al.}{2010}]{2010MNRAS.403...45S}
{Stacy} A.,  {Greif} T.~H.,   {Bromm} V.,  2010, \mn@doi [\mnras]
  {10.1111/j.1365-2966.2009.16113.x}, \href
  {http://adsabs.harvard.edu/abs/2010MNRAS.403...45S} {403, 45}

\bibitem[\protect\citeauthoryear{{Tanaka}, {Perna}  \& {Haiman}}{{Tanaka}
  et~al.}{2012}]{2012MNRAS.425.2974T}
{Tanaka} T.,  {Perna} R.,   {Haiman} Z.,  2012, \mn@doi [\mnras]
  {10.1111/j.1365-2966.2012.21539.x}, \href
  {http://adsabs.harvard.edu/abs/2012MNRAS.425.2974T} {425, 2974}

\bibitem[\protect\citeauthoryear{{Thomas} \& {Zaroubi}}{{Thomas} \&
  {Zaroubi}}{2008}]{thomas08}
{Thomas} R.~M.,  {Zaroubi} S.,  2008, \mn@doi [\mnras]
  {10.1111/j.1365-2966.2007.12767.x}, \href
  {http://adsabs.harvard.edu/abs/2008MNRAS.384.1080T} {384, 1080}

\bibitem[\protect\citeauthoryear{{Thomas} et~al.,}{{Thomas}
  et~al.}{2009}]{Thom09}
{Thomas} R.~M.,  et~al., 2009, \mn@doi [\mnras]
  {10.1111/j.1365-2966.2008.14206.x}, \href
  {http://adsabs.harvard.edu/abs/2009MNRAS.393...32T} {393, 32}

\bibitem[\protect\citeauthoryear{{Tingay} et~al.,}{{Tingay}
  et~al.}{2013}]{tingay13}
{Tingay} S.~J.,  et~al., 2013, \mn@doi [Publications of the Astronomical
  Society of Australia (PASA)] {10.1017/pasa.2012.007}, \href
  {http://adsabs.harvard.edu/abs/2013PASA...30....7T} {30, 7}

\bibitem[\protect\citeauthoryear{{Tozzi}, {Madau}, {Meiksin}  \&
  {Rees}}{{Tozzi} et~al.}{2000}]{2000ApJ...528..597T}
{Tozzi} P.,  {Madau} P.,  {Meiksin} A.,   {Rees} M.~J.,  2000, \mn@doi [\apj]
  {10.1086/308196}, \href {http://adsabs.harvard.edu/abs/2000ApJ...528..597T}
  {528, 597}

\bibitem[\protect\citeauthoryear{{Van Haarlem} et~al.,}{{Van Haarlem}
  et~al.}{2013}]{van13}
{Van Haarlem} M.~P.,  et~al., 2013, \mn@doi [\aap]
  {10.1051/0004-6361/201220873}, \href
  {http://adsabs.harvard.edu/abs/2013A%26A...556A...2V} {556, A2}

\bibitem[\protect\citeauthoryear{{Vanden Berk} et~al.,}{{Vanden Berk}
  et~al.}{2001}]{vanden01}
{Vanden Berk} D.~E.,  et~al., 2001, \mn@doi [\aj] {10.1086/321167}, \href
  {http://adsabs.harvard.edu/abs/2001AJ....122..549V} {122, 549}

\bibitem[\protect\citeauthoryear{{Venemans} et~al.,}{{Venemans}
  et~al.}{2015}]{Venemans15}
{Venemans} B.~P.,  et~al., 2015, \mn@doi [\apjl] {10.1088/2041-8205/801/1/L11},
  \href {http://adsabs.harvard.edu/abs/2015ApJ...801L..11V} {801, L11}

\bibitem[\protect\citeauthoryear{{Vignali}, {Brandt}  \& {Schneider}}{{Vignali}
  et~al.}{2003}]{vignali03}
{Vignali} C.,  {Brandt} W.~N.,   {Schneider} D.~P.,  2003, \mn@doi [\aj]
  {10.1086/345973}, \href {http://adsabs.harvard.edu/abs/2003AJ....125..433V}
  {125, 433}

\bibitem[\protect\citeauthoryear{{Villaescusa-Navarro}, {Viel}, {Datta}  \&
  {Choudhury}}{{Villaescusa-Navarro} et~al.}{2014}]{2014JCAP...09..050V}
{Villaescusa-Navarro} F.,  {Viel} M.,  {Datta} K.~K.,   {Choudhury} T.~R.,
  2014, \mn@doi [\jcap] {10.1088/1475-7516/2014/09/050}, \href
  {http://adsabs.harvard.edu/abs/2014JCAP...09..050V} {9, 050}

\bibitem[\protect\citeauthoryear{{Vonlanthen}, {Semelin}, {Baek}  \&
  {Revaz}}{{Vonlanthen} et~al.}{2011}]{2011A&A...532A..97V}
{Vonlanthen} P.,  {Semelin} B.,  {Baek} S.,   {Revaz} Y.,  2011, \mn@doi [\aap]
  {10.1051/0004-6361/201116811}, \href
  {http://adsabs.harvard.edu/abs/2011A%26A...532A..97V} {532, A97}

\bibitem[\protect\citeauthoryear{{Wang}, {Tegmark}, {Santos}  \& {Knox}}{{Wang}
  et~al.}{2006}]{2006ApJ...650..529W}
{Wang} X.,  {Tegmark} M.,  {Santos} M.~G.,   {Knox} L.,  2006, \mn@doi [\apj]
  {10.1086/506597}, \href {http://adsabs.harvard.edu/abs/2006ApJ...650..529W}
  {650, 529}

\bibitem[\protect\citeauthoryear{{Wise} \& {Abel}}{{Wise} \&
  {Abel}}{2007}]{2007ApJ...665..899W}
{Wise} J.~H.,  {Abel} T.,  2007, \mn@doi [\apj] {10.1086/520036}, \href
  {http://adsabs.harvard.edu/abs/2007ApJ...665..899W} {665, 899}

\bibitem[\protect\citeauthoryear{{Wise}, {Turk}, {Norman}  \& {Abel}}{{Wise}
  et~al.}{2012}]{wise2012}
{Wise} J.~H.,  {Turk} M.~J.,  {Norman} M.~L.,   {Abel} T.,  2012, \mn@doi
  [\apj] {10.1088/0004-637X/745/1/50}, \href
  {http://adsabs.harvard.edu/abs/2012ApJ...745...50W} {745, 50}

\bibitem[\protect\citeauthoryear{{Wise}, {Demchenko}, {Halicek}, {Norman},
  {Turk}, {Abel}  \& {Smith}}{{Wise} et~al.}{2014}]{2014MNRAS.442.2560W}
{Wise} J.~H.,  {Demchenko} V.~G.,  {Halicek} M.~T.,  {Norman} M.~L.,  {Turk}
  M.~J.,  {Abel} T.,   {Smith} B.~D.,  2014, \mn@doi [\mnras]
  {10.1093/mnras/stu979}, \href
  {http://adsabs.harvard.edu/abs/2014MNRAS.442.2560W} {442, 2560}

\bibitem[\protect\citeauthoryear{{Wyithe}, {Geil}  \& {Kim}}{{Wyithe}
  et~al.}{2015}]{2015aska.confE..15W}
{Wyithe} S.,  {Geil} P.,   {Kim} H.,  2015, Advancing Astrophysics with the
  Square Kilometre Array (AASKA14), \href
  {http://adsabs.harvard.edu/abs/2015aska.confE..15W} {p.~15}

\bibitem[\protect\citeauthoryear{{Xu}, {Wise}, {Norman}, {Ahn}  \&
  {O'Shea}}{{Xu} et~al.}{2016}]{2016arXiv160407842X}
{Xu} H.,  {Wise} J.~H.,  {Norman} M.~L.,  {Ahn} K.,   {O'Shea} B.~W.,  2016,
  preprint, \href {http://adsabs.harvard.edu/abs/2016arXiv160407842X} {}
  (\mn@eprint {arXiv} {1604.07842})

\bibitem[\protect\citeauthoryear{{Yajima} \& {Li}}{{Yajima} \&
  {Li}}{2014}]{2014MNRAS.445.3674Y}
{Yajima} H.,  {Li} Y.,  2014, \mn@doi [\mnras] {10.1093/mnras/stu1982}, \href
  {http://adsabs.harvard.edu/abs/2014MNRAS.445.3674Y} {445, 3674}

\bibitem[\protect\citeauthoryear{{Zahn}, {Lidz}, {McQuinn}, {Dutta},
  {Hernquist}, {Zaldarriaga}  \& {Furlanetto}}{{Zahn} et~al.}{2007}]{zahn2007}
{Zahn} O.,  {Lidz} A.,  {McQuinn} M.,  {Dutta} S.,  {Hernquist} L.,
  {Zaldarriaga} M.,   {Furlanetto} S.~R.,  2007, \mn@doi [\apj]
  {10.1086/509597}, \href {http://adsabs.harvard.edu/abs/2007ApJ...654...12Z}
  {654, 12}

\bibitem[\protect\citeauthoryear{{Zaroubi}, {Thomas}, {Sugiyama}  \&
  {Silk}}{{Zaroubi} et~al.}{2007}]{2007MNRAS.375.1269Z}
{Zaroubi} S.,  {Thomas} R.~M.,  {Sugiyama} N.,   {Silk} J.,  2007, \mn@doi
  [\mnras] {10.1111/j.1365-2966.2006.11361.x}, \href
  {http://adsabs.harvard.edu/abs/2007MNRAS.375.1269Z} {375, 1269}

\bibitem[\protect\citeauthoryear{{Zaroubi} et~al.,}{{Zaroubi}
  et~al.}{2012}]{2012MNRAS.425.2964Z}
{Zaroubi} S.,  et~al., 2012, \mn@doi [\mnras]
  {10.1111/j.1365-2966.2012.21500.x}, \href
  {http://adsabs.harvard.edu/abs/2012MNRAS.425.2964Z} {425, 2964}

\makeatother
\end{thebibliography}

\appendix

%%%%%%%%%%%%%%%
\section{Removing the foregrounds using a filter}
\label{filter_fg}

\begin{figure*}
\begin{center}
\includegraphics[scale=0.7]{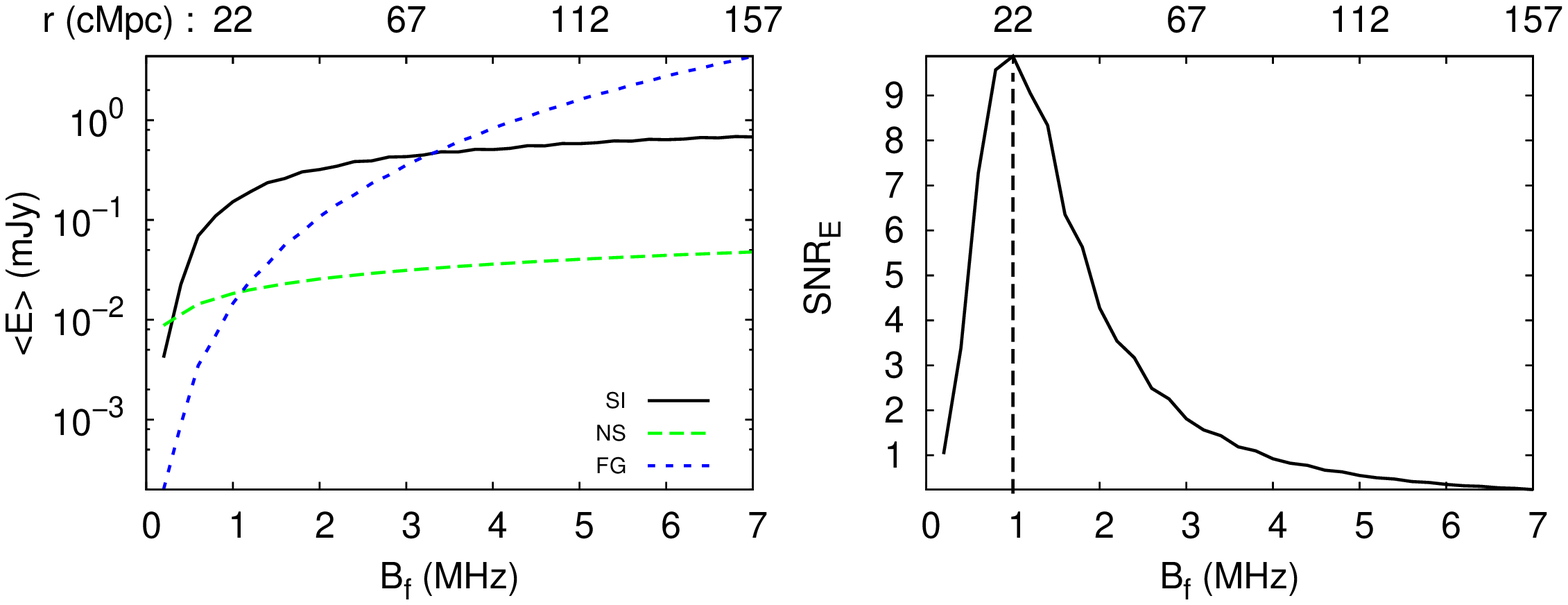}
    \caption{Left-hand panel: The estimator of the signal using the filter $S_f$ and the corresponding error from the system noise and the foregrounds as a function of the parameter $B_f$. Right-hand panel: The signal to noise ratio as a function of the parameter $B_f$. The top labels of the panels show the spatial scales corresponds to $B_f$.}
   \label{image_p4filter}
\end{center}
\end{figure*}

In this section, we present a different approach to remove the foregrounds using a suitable filter rather than subtracting the foregrounds using some subtraction method as explained earlier. We showed in Paper I that  it is possible to detect the signal by summing up all the visibility contributions from different baselines and frequency channels in the foreground-free scenario. Even in the presence of the foregrounds, the signal is detectable using suitable filters which can decrease the foregrounds contributions below the signal. While the previous work was done using the analytical form of the signal from the source and for a somewhat idealized baseline distribution, here we repeat the calculation using the simulated signal and recently published baseline distribution of the SKA1-low.

 The filter used introduced in Paper I does not depend on any prior information on the signal and it only uses the fact that the foregrounds have smooth frequency dependence. The details of the method of subtracting foregrounds using filters can be found in previous works like \citet{kanan2007MNRAS.382..809D, ghara15c}. Here we briefly describe the method.

We define the estimator $\hat{E}$ as
\begin{equation}
\hat{E}= A_{\rm NS} (\Delta U)^2 \Delta {\nu}_c \sum_{a, b}  ~V(\vec{U_a}, {\nu}_b) ~S^{\star}_f(\vec{U_a}, {\nu}_b) ~{n}_{\rm B}(\vec{U_a}, {\nu}_b),
\label{equ_est}
\end{equation} 
where $\Delta U$ is the grid resolution in the  baseline distribution and the quantity $S_f(\vec{U}, \nu)$ represents the filter. The sum is over all possible baselines $a$ and frequency channels $b$. The normalization constant $A_{\rm NS}$ is given by
\begin{equation}
A_{\rm NS}^{-1} = (\Delta U)^2 \Delta {\nu}_c \sum_{a, b} {n}_{\rm B}(\vec{U_a}, {\nu}_b) = N_{\rm B} B_{\nu},
\label{equ_ANS}
\end{equation}
where $N_{\rm B}$ is the total number of baselines used in the study.

The system noise and the foregrounds are expected to be random numbers with zero mean. Thus, the expectation value of the estimator is expected to be, 
\begin{equation}
\left< \hat{E}\right>=A_{\rm NS} (\Delta U)^2 \Delta {\nu}_c \sum_{a, b}  ~S(\vec{U_a}, {\nu}_b) ~S^{\star}_f(\vec{U_a}, {\nu}_b) ~{n}_{\rm B}(\vec{U_a}, {\nu}_b).
\label{equ_exp_est}
\end{equation}
The associated errors from the system noise can be written as \citep{kanan2007MNRAS.382..809D},
\begin{eqnarray}
\left< (\Delta\hat{E})^2\right>_{\rm NS} \!\!\!\! & = & \!\!\!\! {{\sigma}^2_N} A_{\rm NS} (\Delta U)^2 \Delta {\nu}_c  \nonumber\\  
&\times& \!\!\!\!  \sum_{a, b}  ~|S_f(\vec{U_a}, \nu_b)|^2 ~{n}_{\rm B}(\vec{U_a}, \nu_b),
\label{equ_exp_est_NS}
\end{eqnarray}
where the quantity $\sigma_N$ is given by,
\begin{equation}
 \sigma_N = \frac{\sqrt{2}~ k_B T_{\rm sys}} {A_{\rm eff}~\sqrt{t_{\rm obs}~B_{\nu}~N_{\rm base}}}.
\label{equ_sigN}
\end{equation}
The error contribution from the foregrounds is
\begin{eqnarray}
\left< (\Delta\hat{E})^2\right>_{\rm FG} \!\!\!\! & = & \!\!\!\! A^2_{\rm NS} (\Delta U \Delta {\nu}_c)^2 \sum_{a, b, q}   \left( \frac{2 k_B}{c^2} \right)^2 \left({\nu}_b {\nu}_q \right)^2 \nonumber\\
&\times& \!\!\!\! {n}_{\rm B}(\vec{U_a}, {\nu}_b) ~{n}_{\rm B}(\vec{U_a}, {\nu}_q) ~C_{2\pi U_a}({\nu}_b, {\nu}_q) \nonumber\\
&\times& \!\!\!\! S^{\star}_f(\vec{U_a}, {\nu}_b)S_f(\vec{U_a}, {\nu}_q),
\label{equ_fore_NS}
\end{eqnarray}
where $C_l(\vec{U}, \nu_1, \nu_2)$ represent the multi-frequency angular power spectrum of the total foregrounds. The signal to noise ratio in this method is
\begin{equation}
{\rm SNR} = \frac{\left< \hat{E}\right>}{\sqrt{\left< (\Delta\hat{E})^2\right>_{\rm NS} + \left< (\Delta\hat{E})^2\right>_{\rm FG}}}.
\label{snr_est}
\end{equation}

The form of the filter $S_f$, as defined in Paper I, is taken to be
\begin{eqnarray}
S_f(\vec{U}, \nu) \!\!\!\! & = & \!\!\!\! \left(\frac{\nu}{{\nu}_c} \right)^2 \left[  S_{T}(\vec{U}, \nu, B_{f}) - \frac{\Theta(1-\vert\nu - {\nu}_c\vert/B^{'})}{B^{'}}\right. \nonumber\\
&\times & \!\!\!\!\left. {\int}_{{\nu}_c-B^{'}/2}^{{\nu}_c+B^{'}/2}S_{T}(\vec{U}, {\nu}^{'}, B_{f}) ~{\rm d}{\nu}^{'}  \right], 
\label{equ_filter}
\end{eqnarray}
where 
\begin{eqnarray}
S_{T}(\vec{U}, \nu, B_f) \!\!\!\! & = & \!\!\!\! 0 \mbox{ if } |\nu - {\nu}_c|>\frac{B_f}{2} \nonumber\\
&=& \!\!\!\!  -1  \mbox{ if } |\nu - {\nu}_c|\leq \frac{B_f}{2}.
\label{equ_filter1}
\end{eqnarray}
We choose $B^{'}=2 B_f$ if $B^{'} \leq B_{\nu}$, else $B^{'}= B_{\nu}$. The form of the filter $S_f$ depends on the bandpass filter $S_T(\vec{U}, \nu, B_f)$  of width $B_f$. One can use other more sophisticated filters like the match filter \citep[see, e.g.,][]{kanan2007MNRAS.382..809D} to obtain higher SNR. However, for those filters, one usually requires some prior information about the expected signal.

The main result of the filtering method is shown in Figure \ref{image_p4filter}. The left-hand panel shows the signal estimator and the corresponding errors from the system noise and the foregrounds as a function of the bandpass width $B_f$. We find that it is possible to reduce the foregrounds contribution below the signal level using suitable bandpass width. The  maximum SNR, as shown in the  right-hand panel of Figure \ref{image_p4filter}, is achieved for $B_f \sim 1 $ MHz, and the peak SNR turns out to be $\sim 10$. The width of the filter that provides the maximum SNR, in fact, corresponds to the size of the absorption region around the source.

%\bsp	% typesetting comment
\label{lastpage}
\end{document}